\documentclass[12pt]{article}
\usepackage[utf8]{inputenc}
\usepackage{float}
\usepackage[small,bf]{caption}
\usepackage{fullpage,graphicx,psfrag,url,ar,rotating,hyperref}
\usepackage{amsmath,amssymb,amsthm,enumitem,subcaption,appendix}
\usepackage{makecell}
\usepackage{multirow}

\newcommand{\card}{\mathbf{card}}

\newcommand{\reals}{\mathbf{R}}

\title{Simultaneous Reduction of Number of Spots and Energy Layers in Intensity Modulated Proton Therapy for Rapid Spot Scanning Delivery}
\author{Anqi Fu \and Vicki T. Taasti \and Masoud Zarepisheh}
\date{\today}

\begin{document}
	
\maketitle

\begin{abstract}
	\noindent {\bf Background:} Reducing proton treatment time improves patient comfort and decreases the risk of error from intra-fractional motion, but must be balanced against clinical goals and treatment plan quality.\\
	{\bf Purpose:} To improve the delivery efficiency of spot scanning proton therapy by simultaneously reducing the number of spots and energy layers using the reweighted $l_1$ regularization method.\\
	{\bf Methods:} We formulated the proton treatment planning problem as a convex optimization problem with a cost function consisting of a dosimetric plan quality term plus a weighted $l_1$ regularization term. We iteratively solved this problem and adaptively updated the regularization weights to promote the sparsity of both the spots and energy layers. The proposed algorithm was tested on four head-and-neck cancer patients, and its performance, in terms of reducing the number of spots and energy layers, was compared with existing standard $l_1$ and group $l_2$ regularization methods. We also compared the effectiveness of the three methods ($l_1$, group $l_2$, and reweighted $l_1$) at improving plan delivery efficiency without compromising dosimetric plan quality by constructing each of their Pareto surfaces charting the trade-off between plan delivery and plan quality.\\
	{\bf Results:} The reweighted $l_1$ regularization method reduced the number of spots and energy layers by an average over all patients of $40\%$ and $35\%$, respectively, with an insignificant cost to dosimetric plan quality. From the Pareto surfaces, it is clear that reweighted $l_1$ provided a better trade-off between plan delivery efficiency and dosimetric plan quality than standard $l_1$ or group $l_2$ regularization, requiring the lowest cost to quality to achieve any given level of delivery efficiency.\\
	{\bf Conclusions:} Reweighted $l_1$ regularization is a powerful method for simultaneously promoting the sparsity of spots and energy layers at a small cost to dosimetric plan quality. This sparsity reduces the time required for spot scanning and energy layer switching, thereby improving the delivery efficiency of proton plans.\\
\end{abstract}

\section{Introduction}
\label{sec:introduction}

Intensity modulated proton therapy (IMPT) is typically delivered via the pencil beam scanning technique. The patient is irradiated by a sequence of proton spots, arranged laterally to cover the treatment volume, where the depth of penetration of each spot is determined by its energy layer. During IMPT, protons are transmitted spot-by-spot within every energy layer, and layer-by-layer within every beam over a set of fixed-angle beams. The total plan delivery time is roughly equal to the total switching time between beams (gantry rotation plus beam setup time), switching time between energy layers, travel time between spots, and dose delivery time at each spot \cite{GaoLin:2020,ZhangShenGao:2022}.

In this study, we seek to reduce IMPT delivery time by reducing the number of proton spots and energy layers. A shorter treatment time is desirable because it improves patient comfort, increases patient throughput (hence lowering treatment costs), and decreases the risk of error due to intra-fractional motion \cite{LiZhuZhang:2015,SuzukiLee:2016,MahChenChon:2020}. Simultaneously, we want to ensure clinical goals are met and the quality of the treatment plan remains uncompromised. This trade-off between delivery time and plan quality has been the subject of abundant research.

One thread of research focuses on greedy algorithms for energy layer assignment. These algorithms combine a variety of techniques for control point sampling, energy layer distribution, energy layer filtration, and spot optimization \cite{SPArc:2016}. The goal is to reduce energy layer switching time by pruning the number of energy layers and sequencing them so layer switches only occur from low-to-high energy level \cite{SPArc:2020,EngwallFredriksson:2022}.

Another strand of research takes a structured optimization-based approach. For example, M{\"u}ller and Wilkens (2016) \cite{MullerWilkens:2016} directly minimize the sum of spot intensities as part of a prioritized optimization routine. Other authors formulate the proton treatment delivery problem as a mixed-integer program (MIP), where each energy layer \cite{CaoZhang:2014} or path between layers \cite{ProtonOptSurvey,WuyckensLee:2022} is associated with a binary indicator variable. Their objective is to minimize the dose fidelity (e.g., over/underdose to the target) plus some penalty or constraints on the energy layers, which promote a lower switching time. Although mathematically elegant, these MIPs are computationally difficult to solve, as they scale poorly with the number of energy layers due to the combinatorial nature of the problem.

To avoid this issue, researchers turned their attention to continuous optimization models. A proton treatment planning problem in this category only contains continuous variables, like spot intensities and doses. Typically, the objective includes a regularization function that is selected to encourage sparsity (i.e., more zero elements) in the spots and energy layers. The regularizer applies a penalty to the total spot intensity within each layer group. A variety of options have been proposed for this penalty function: logarithm \cite{WaterHoogeman:2015,WaterLomax:2020}, $l_{2,1/2}$ norm \cite{GuSheng:2020}, and $l_2$ norm \cite{JensenAlber:2018,LinClasieGao:2019,GaoLin:2020,ProtonOptSurvey}. The last of these is of particular interest because it is convex and widely used in statistics for promoting group sparsity; the associated regularizer is known as the group lasso \cite{YuanLin:2006,MeierGeerBuhlmann:2008,LimHastie:2015,IvanoffRivoirard:2016}.

The standard lasso (i.e., $l_1$ norm penalty) promotes sparsity of the spot intensities, but does not directly penalize the energy layers. The group lasso (i.e., group $l_2$ norm penalty) promotes sparsity of the energy layers, but actually {\em increases} the number of nonzero spots. Since IMPT delivery time depends on both the number of spots and energy layers \cite{GaoLin:2020}, neither of these regularizers is ideal. In this paper, we propose a new regularization method that simultaneously reduces the number of nonzero spots and energy layers, while upholding treatment plan quality. Our reweighted $l_1$ method combines the $l_1$ penalty from standard lasso with a weighting mechanism that differentiates between the spots of different energy layers, similar to the group lasso. We test the proposed method on four head-and-neck cancer patients and demonstrate its ability to 1) reduce the number of spots and energy layers simultaneously, and 2) provide a better trade-off between dosimetric plan quality and plan delivery efficiency than existing regularization methods (i.e., standard $l_1$ and group lasso).



\section{Methods and materials}
\label{sec:methods}

\subsection{Problem formulation}
\label{sec:problem}


We discretize the patient's body into $m$ voxels and the proton beams into $n$ spots. For each spot $j \in \{1,\ldots,n\}$, we calculate the radiation dose delivered by a unit intensity of that spot to voxel $i \in \{1,\ldots,m\}$ and call this value $A_{ij}$. The dose influence matrix is then $A \in \reals_+^{m \times n}$, where its rows correspond to the voxels and its columns to the spots. Let $p \in \reals_+^m$ be the prescription vector, i.e., $p_i$ equals the physician-prescribed dose if $i$ is a target voxel and zero otherwise. We concatenate the spot intensities of all beams and all energy layers within each beam, denoting this collective as the vector $x \in \reals^n$. The typical treatment planning problem seeks a vector of spot intensities $x$ that minimizes the deviation of the delivered dose, $d = Ax$, from the prescription $p$. This deviation can be decomposed into a penalty on the overdose, $\overline d = (d - p)_+ = \max(d - p,0)$, and the underdose, $\underline d = (d - p)_- = -\min(d - p,0)$, which we combine to form the {\em cost function}
\begin{equation}
	\label{eq:cost}
	f(\overline d, \underline d) = \sum_{i=1}^m \overline w_i \overline d_i^2 + \underline w_i \underline d_i^2,
\end{equation}
where $\overline w, \underline w \in \reals_+^n$ are penalty parameters that determine the relative importance of the over/underdose to the treatment plan. (Note the underdose is ignored for non-target voxels $i$ because $p_i = 0$).

Dose constraints are defined for each anatomical structure. For a given structure $s$, let $A^s \in \reals_+^{m_s \times n}$ be the row slice of $A$ containing only the rows of the $m_s$ voxels in $s$. A maximum dose constraint takes the form of $A^sx \leq d_s^{\textrm{\tiny max}}$, where $d_s^{\textrm{\tiny max}}$ is an upper bound. Similarly, a mean dose constraint is of the form $\frac{1}{m_s}\mathbf{1}^TA^sx \leq d_s^{\textrm{\tiny mean}}$.
By stacking the constraint matrices/vectors for all $S$ structures, we can represent the set of dose constraints as a single linear inequality $Bx \leq c$, where $B = [A^1, \frac{1}{m_1}\mathbf{1}^TA^1, \ldots, A^S, \frac{1}{m_S}\mathbf{1}^TA^S]$ and $c = [d_1^{\textrm{\tiny max}}, d_1^{\textrm{\tiny mean}}, \ldots, d_S^{\textrm{\tiny max}}, d_S^{\textrm{\tiny mean}}]$. Then, our treatment planning problem is
\begin{equation}
	\label{prob:unreg_ncvx}
	\begin{array}{ll}
		\mbox{minimize} & f(\overline d, \underline d) \\
		\mbox{subject to} & \overline d = (Ax - p)_+, \quad \underline d = (Ax - p)_-, \quad Bx \leq c \\
		& x \geq 0, \quad \overline d \geq 0, \quad \underline d \geq 0
	\end{array}
\end{equation}
with variables $x \in \reals^n, \overline d \in \reals^m$, and $\underline d \in \reals^m$. Since the objective function $f$ is monotonically increasing in $\overline d$ and $\underline d$ over the nonnegative reals, we can write this problem equivalently as
\begin{equation}
	\label{prob:unreg}
	\begin{array}{ll}
		\mbox{minimize} & f(\overline d, \underline d) \\
		\mbox{subject to} & Ax - \overline d + \underline d = p, \quad Bx \leq c \\
		& x \geq 0, \quad \overline d \geq 0, \quad \underline d \geq 0.
	\end{array}
\end{equation}
(The derivation is provided in appendix \ref{app:prob_simp}). Problem \ref{prob:unreg} is a convex quadratic program (QP), hence can be solved using standard convex methods, e.g., the alternating direction method of multipliers (ADMM) \cite{ADMM,robust_proton} or interior-point methods \cite{Wright:1997,Gorissen:2022}. The reader is referred to S. Boyd and L. Vandenberghe (2004) \cite{BoydVandenberghe:2004} and J. Nocedal and S. J. Wright (2006) \cite{NocedalWright:2006} for a thorough discussion of convex optimization.

\subsection{Common regularizers}
\label{sec:regularizers}

The cost function defined in \ref{eq:cost} focuses solely on the difference of the delivered dose from the prescription, i.e., the dosimetric plan quality. However, in our treatment scenario, we are also interested in reducing the dose delivery time, i.e., increasing the plan delivery efficiency. The delivery time is positively correlated with the number of nonzero spots (spot scanning rate) and nonzero energy layers (energy switching time) \cite{GaoLin:2020,PoulsenLangen:2018}. Thus, we want to augment the objective of problem \ref{prob:unreg} with a {\em regularization function} $r:\reals^n \rightarrow \reals$, which penalizes the spot vector $x$ in a way that reduces the number of nonzero spots/layers, while maintaining high plan quality.
The regularized treatment planning problem is 
\begin{equation}
\label{prob:reg}
	\begin{array}{ll}
		\mbox{minimize} & f(\overline d, \underline d) + \lambda r(x) \\
		\mbox{subject to} & Ax - \overline d + \underline d = p, \quad Bx \leq c \\
		& x \geq 0, \quad \overline d \geq 0, \quad \underline d \geq 0
	\end{array}
\end{equation}
with respect to $x, \overline d$, and $\underline d$. Here we have introduced a  regularization weight $\lambda \geq 0$ to balance the trade-off between dosimetric plan quality, represented by the cost $f(\overline d, \underline d)$, and plan delivery efficiency, as captured by the regularization term $r(x)$. A larger value of $\lambda$ places more importance on efficiency.

In the following subsections, we review a few regularization functions that have been suggested in the literature. Let $\mathcal{J} = \{1,\ldots,n\}$ and $\mathcal{G} = \{\mathcal{J}_1,\ldots,\mathcal{J}_G\}$ be a set of subsets of $\mathcal{J}$, where each $\mathcal{J}_g \subseteq \mathcal{J}$ has exactly $n_g \leq n$ elements. Specifically in our setting, $\mathcal{G}$ represents a partition of $n$ spots into $G$ energy layers with $\mathcal{J}_g$ containing the indices of the $n_g$ spots in layer $g$.

\subsubsection{$l_0$ regularizer}
One method of reducing the delivery time is to directly penalize the number of nonzero spots. This can be accomplished via the $l_0$ regularizer
\begin{equation}
\label{eq:l0_std}
	r_0(x) = \|x\|_0 = \card(\{j: x_j \neq 0\}),
\end{equation}
which we have defined as the number of nonzero elements in $x$. 
(Here $\card(A)$ denotes the cardinality of set $A$). 
Unfortunately, the $l_0$ regularization function is computationally expensive to implement. To solve problem \ref{prob:reg} with $r = r_0$, we would need to solve a series of large mixed-integer programs in order to determine the optimal subset of nonzero spots out of all possible combinations from $\mathcal{J}$ \cite{BertsimasMazumder:2016}. As the number of spots $n$ is typically very large (on the order of $10^3$ to $10^4$), this quickly becomes computationally intractable.

Another option is to apply the $l_0$ regularizer to the energy layers:
\begin{equation}
\label{eq:l0_grp}
	\tilde r_0(x) = \left\|\left[\sum_{j \in \mathcal{J}_1} x_j, \ldots, \sum_{j \in \mathcal{J}_G} x_j\right]\right\|_0 = \card\left(\left\{g: \sum_{j \in \mathcal{I}_g} x_j \neq 0 \right\}\right).
\end{equation}
In this case, $\tilde r_0$ returns the number of nonzero energy layers, where a layer $g$ is zero if and only if all its spots are zero, i.e., $\sum_{j \in \mathcal{J}_g} x_j = 0$. The associated combinatorial problem or MIP simplifies to finding the optimal subset of nonzero layers, which is more manageable since $G$ is typically on the order of $10^2$. Cao et al. (2014) \cite{CaoZhang:2014} developed an iterative method to solve an approximation of this MIP and were able to reduce the number of proton energies in their IMPT plan, while satisfying certain dosimetric criteria. Nevertheless, as combinatorial optimization is still expensive, we turn our attention to a different regularization function.

\subsubsection{$l_1$ regularizer}
A common approximation of the $l_0$ regularizer is the $l_1$ norm. Define the $l_1$ regularization function to be
\begin{equation}
\label{eq:l1_std}
	r_1(x) = \|x\|_1 = \sum_{j=1}^n |x_j|.
\end{equation}
This function is closed, convex, and continuous. When used as a regularizer in problem \ref{prob:reg}, it produces a convex optimization problem -- a form of the lasso problem -- that promotes sparsity in the solution vector $x$ \cite{Tibshirani:1996}.
The lasso problem is well-studied in the literature \cite{L1Survey,LassoBook}, and various methods have been developed to solve it quickly and efficiently \cite{ParkHastie:2007,SchmidtRosales:2007,WuLange:2008,ShiSajda:2010}.

One downside of the $l_1$ regularizer is that it does not differentiate between energy layers and thus is insensitive to the number of layers: since the spot vector is nonnegative, the absolute value of its elements $|x_j| = x_j$, and any sum over energy layers decouples into the sum over all spots $\sum_{g=1}^G \sum_{j \in \mathcal{J}_g} |x_j| = \sum_{j=1}^n x_j$. As energy layer switching time is typically longer than spot delivery or travel time, $l_1$ regularization is not the most effective method for improving plan delivery efficiency.

\subsubsection{Group $l_2$ regularizer}
The group $l_2$ regularizer, also known as the group lasso, provides an alternative method for efficiently implementing group penalties. This regularization function is defined as
\begin{equation}
\label{eq:l2_grp}
	r_2(x) = \sum_{g=1}^G \frac{1}{\sqrt{n_g}} \|\{x_j: j \in \mathcal{J}_g\}\|_2 = \sum_{g=1}^G \sqrt{\frac{1}{n_g} \sum_{j \in \mathcal{J}_g} x_j^2}.
\end{equation}
It is the sum of the $l_2$ norm of the vector corresponding to each group (i.e., energy layer), weighted by the reciprocal of the square root of the total number of group elements \cite{YuanLin:2006}. (The weights $\frac{1}{\sqrt{n_g}}$ may differ across applications; see N. Simon and R. Tibshirani (2012) \cite{GroupLassoWeights} for alternatives). Group lasso has been widely researched in the context of statistical analysis and regression \cite{MeierGeerBuhlmann:2008,JacobVert:2009,LimHastie:2015,IvanoffRivoirard:2016,MaSongHuang:2007}, and many algorithms exist for solving the associated optimization problem effectively \cite{Rakoto:2011,QinGoldfarb:2013,YangZou:2015}. Jensen et al. (2018) \cite{JensenAlber:2018} employed a version of the group lasso to perform adaptive IMPT energy layer optimization.

In our proton treatment delivery scenario, the group lasso is capable of differentiating between energy layers, and thus is a good regularizer for reducing the number of nonzero layers. However, it also tends to {\em increase} the number of nonzero spots within the active layers, as the quadratic penalty term in equation \ref{eq:l2_grp} mostly ignores small spot intensities (square of a small $x_j$ is near zero).
This failure to generate sparsity in the spot vector due to the characteristics of the $l_2$ norm make it an inadequate regularizer for our purposes.



\subsection{Reweighted $l_1$ method}
\label{sec:l1_reweight}

As we have discussed, the $l_1$ regularization function promotes sparsity of the spots, but not the energy layers. The group $l_2$ regularization function promotes sparsity of the layers, but not the spots -- indeed, it tends to produce dense spot vectors due to the $l_2$ norm. In this section, we introduce the reweighted $l_1$ regularization method, which promotes sparsity in both the spots and the energy layers.



The reweighted $l_1$ method assigns a weight to every spot based on its intensity and energy layer. The weights are chosen to counteract the intensity of each layer, so that all spots, regardless of intensity, contribute roughly equally to the total regularization penalty. This is done to imitate the ``ideal'' group $l_0$ regularizer (equation \ref{eq:l0_grp}), which counts every nonzero layer as one unit (due to $\card$) regardless of intensity.

Formally, we define the {\em weighted group $l_1$ regularizer}
\begin{equation}
\label{eq:l1_weight}
	r_3(x;\beta) = \sum_{g=1}^G \beta_g \sum_{j \in \mathcal{J}_g} |x_j|
\end{equation}
with weight parameters $\beta_g \in \reals_+ \cup \{+\infty\}$ for $g = 1,\ldots,G$. An intuitive way to set $\beta_g$ is to make it inversely proportional to the optimal total intensity of energy layer $g$, i.e.,
\[
	\beta_g = \begin{cases}
		\frac{1}{\sum_{j \in \mathcal{J}_g} x_j^{\star}} & \sum_{j \in \mathcal{J}_g} x_j^{\star} \neq 0 \\
		+\infty & \sum_{j \in \mathcal{J}_g} x_j^{\star} = 0 
	\end{cases},
\]
where $x^{\star} \in \reals_+^n$ is an optimal spot vector. 
However, we do not know $x^{\star}$ beforehand. We will approximate this weighting scheme iteratively using the reweighted $l_1$ method.

The reweighted $l_1$ method is a type of majorization-minimization (MM) algorithm, which solves an optimization problem by iteratively minimizing a surrogate function that majorizes the actual objective function. MM algorithms have a rich history in the literature \cite{FigueiredoNowak:2007,SchifanoWells:2010,Mairal:2015,SunPalomar:2017}, and reweighted $l_1$ in particular has been used to solve problems in portfolio optimization \cite{LoboFazelBoyd:2007}, matrix rank minimization \cite{FazelThesis,FazelHindiBoyd:2003}, and sparse signal recovery \cite{CandesWakinBoyd:2008}. Research has shown that it is fast and robust, outperforming standard $l_1$ regularization in a variety of settings.

We now describe the reweighted $l_1$ method in our treatment planning setting. Let the initial weights $\beta^{(1)} = \mathbf{1}$. At each iteration $k = 1,2,\ldots$,
\begin{enumerate}
	\item Set $x^{(k)}$ to a solution of
	\begin{equation}
		\label{prob:l1_reweight}
		\begin{array}{ll}
			\mbox{minimize} & f(\overline d, \underline d) + \lambda r_3(x; \beta^{(k)}) \\
			\mbox{subject to} & Ax - \overline d + \underline d = p, \quad Bx \leq c \\
			& x \geq 0, \quad \overline d \geq 0, \quad \underline d \geq 0.
		\end{array}
	\end{equation}
	\item Compute the total intensity of each energy layer $e_g^{(k)} = \sum_{j \in \mathcal{J}_g} x_j^{(k)}$.
	\item Lower threshold the solution
	\[
		\tilde e_g^{(k)} = \max(e_g^{(k)}, \epsilon^{(k)}), \quad g = 1,\ldots,G,
	\]
	where $\epsilon^{(k)} = \delta \max_{g'} e_{g'}^{(k)}$ for some small $\delta \in (0,1)$.
	\item Update the weights. First, compute the standardized reciprocals
	\begin{equation}
		\label{eq:l1_reweight_scale_1}
		\alpha_g^{(k)} = \left(\frac{1}{\tilde e_g^{(k)}}\right) \bigg / \left(\sum_{g' = 1}^G \frac{1}{\tilde e_{g'}^{(k)}}\right), \quad g = 1,\ldots,G.
	\end{equation}
	Then, calculate the scaling term
	\begin{equation}
		\label{eq:l1_reweight_scale_2}
		\mu^{(k)} = \sum_{g=1}^G \tilde e_g^{(k)} \bigg / \left( \sum_{g=1}^G \alpha_g^{(k)} \tilde e_g^{(k)} \right).
	\end{equation}
	The new weights are $\beta_g^{(k+1)} = \mu^{(k)} \alpha_g^{(k)}$.
	\item Terminate on convergence of the objective, or when $k$ reaches a user-defined maximum number of iterations $K$.
\end{enumerate}
Step 3 was introduced to ensure stability of the algorithm, so that a zero energy layer estimate $e_g^{(k)} = 0$ would not preclude the subsequent estimate $e_g^{(k+1)}$ from being nonzero. In our computational experiments, we found that a threshold fraction of $\delta = 0.01$ produced good results. Step 4 was added to ensure the $l_1$ regularization term ($r_1(x)$) and the reweighted $l_1$ term ($r_3(x)$) contribute a similar amount to the objective. To this end, the reweighted $l_1$ term is scaled by $\mu^{(k)}$ so that $r_1(x^{(k)}) = r_3(x^{(k)})$.

The reweighted $l_1$ method has a number of advantages over the other regularizers we have reviewed here. It encourages sparsity in energy layers by properly grouping the $l_1$ penalty. It spreads this penalty evenly across all energy layers using its weighting scheme, rather than prioritizing those spots with large magnitudes. Finally, it is easy to implement: each iteration of the algorithm only requires that we solve a simple convex problem with $l_1$ regularization, which can be done efficiently using many off-the-shelf solvers. Moreover, the number of reweighting iterations needed in practice is typically very low, with most of the improvement coming from the first two to three iterations, so its computational cost is overall low. As we will see in the next section, reweighted $l_1$ outperforms regular $l_1$ and group $l_2$ penalties in sparsifying spots/layers.

\subsection{Patient population and computational framework}
\label{sec:patients}

We compared the reweighted $l_1$ method with standard $l_1$ and group $l_2$ regularization on four head-and-neck cancer patient cases from The Cancer Imaging Archive (TCIA) \cite{TCIA:paper,TCIA:data}. For each patient, we created the dose influence matrix using the proton pencil beam calculation engine in the open-source package MatRad \cite{MatRad,MatRad:Int}, assuming a relative biological effectiveness (RBE) of $1.1$. The proton spots were situated on a rectangular grid with a spot spacing of $5$ mm, and the grid covered the entire PTV plus $1$ mm out from its perimeter. The voxel resolution and dose grid resolution were both $0.98$ mm $\times$ $0.98$ mm $\times$ $2$ mm. Every patient plan was generated using two co-planar beams with beam angles selected using a Bayesian optimization algorithm \cite{TaastiHongZarepisheh:2020}. Table \ref{table:patient_info} provides more details.

\begin{table}
	\centering
	\begin{tabular}{l|rrrr}
		\hline \hline
		& \multicolumn{4}{c}{Patient} \\
		\cline{2-5}
		& 1 & 2 & 3 & 4 \\
		\hline
		Beam configuration & $40^{\circ}, 90^{\circ}$ & $74^{\circ}, 285^{\circ}$ & $75^{\circ}, 130^{\circ}$ & $220^{\circ}, 290^{\circ}$ \\ 
		PTV volume (cm$^3$) & 162.7 & 169.7 & 129.9 & 12.4 \\
		Number of voxels ($m$) & 87012 & 117907 & 110869 & 50728 \\
		Number of spots ($n$) & 6378 & 7011 & 5257 & 713 \\
		Number of energy layers ($G$) & 56 & 70 & 62 & 38 \\
		\hline \hline
	\end{tabular}
	\caption{Head-and-neck cancer patient information.}
	\label{table:patient_info}
\end{table}

Each patient had a single planning target volume (PTV) that was prescribed a dose of $p = 70$ Gy delivered in $35$ fractions of $2$ Gy per fraction. The mean/max dose bounds for important structures are given in Table \ref{table:dose_constrs}. We manually adjusted the weights in cost function \ref{eq:cost}, without exhaustive search, to obtain reasonable treatment plans. In the majority of cases, $\overline w_i = 1, \underline w_i = 10$ for PTV voxels $i$ and $\overline w_i = 10^{-3}$ for all other voxels $i$ provided the best results.

\begin{table}
	\centering
	\begin{tabular}{l|rr}
		\hline \hline
		& \multicolumn{2}{c}{Dose Bound (Gy)} \\
		\cline{2-3}
		& $d^{\textrm{\tiny max}}$ & $d^{\textrm{\tiny mean}}$ \\
		\hline
		PTV           & 84   & N/A \\
		Left Parotid  & 73.5 & 26 \\
		Right Parotid & 73.5 & 26 \\
		Mandible      & 70   & N/A \\
		Spinal Cord   & 45   & N/A \\
		Constrictors  & N/A  & 40 \\
		Brainstem     & 54   & N/A \\
		\hline \hline
	\end{tabular}
	\caption{Dose constraints for each structure. N/A means the max/mean dose was unbounded, i.e., $d^{\textrm{\tiny max}} = +\infty$ or $d^{\textrm{\tiny mean}} = +\infty$.}
	\label{table:dose_constrs}
\end{table}

We implemented the standard $l_1$, group $l_2$, and reweighted $l_1$ based treatment planning methods in Python using CVXPY \cite{DiamondBoyd:2016,AgrawalBoyd:2018,cvxpy} and solved the associated optimization problems with MOSEK \cite{mosek}. All computational processes were executed on a $64$-bit PC with an AMD Ryzen 9 3900X CPU @ $3.80$ GHz/ $12$ cores and $128$ GB RAM. For reweighted $l_1$, we ran the algorithm for $K = 3$ iterations due to the diminishing benefits of more iterations.

To facilitate comparisons, we scaled the group $l_2$ regularizer so it lay in the same range as the standard $l_1$ regularizer. First, we solved problem \ref{prob:reg} with the standard $l_1$ regularizer (equation \ref{eq:l1_std}) and $\lambda = 1$. Let us call this solution $x^{(1)}$. Then, we computed a scaling term $\eta > 0$ such that $\eta r_2(x^{(1)}) = r_1(x^{(1)})$. When we ran the group $l_2$ method, we used the scaled regularization function $\tilde r_2(x;\eta) := \eta r_2(x)$ as the regularizer $r(x)$ in problem \ref{prob:reg}. This allowed us to obtain better spot/energy layer comparison plots between the standard $l_1$ and group $l_2$ methods. As pointed out earlier, the reweighted $l_1$ method is similarly scaled via step 4 of the algorithm.

After each method finished, we trimmed the optimal spot vector $x^{\star}$ further to increase sparsity. First, we zeroed out all elements $x_j^{\star}$ that fell below a fraction $\gamma \in (0,1)$ of the maximum spot intensity, i.e., we set $x_j^{\star} = 0$ if $x_j^{\star} < \gamma \max_{j'} x_{j'}^{\star}$. We then zeroed out all energy layers of the resulting $\tilde x^{\star}$ that fell below the same fraction of the maximum layer intensity: for each $g \in \{1,\ldots,G\}$, we set $\tilde x_j^{\star} = 0$ for all $j \in \mathcal{J}_g$ if $\sum_{j' \in \mathcal{J}_g} \tilde x_{j'}^{\star} < \gamma \max_{g'} \sum_{j' \in \mathcal{J}_{g'}} x_{j'}^{\star}$. A choice of $\gamma = 0.01$ provided a reasonable trade-off between sparsity and dose coverage in our computational experiments.

\section{Results}
\label{sec:results}

\subsection{Simultaneous reduction of spots and energy layers}
\label{sec:res_spots}

\begin{figure}[H]
	\centering
	\includegraphics[width=0.9\linewidth]{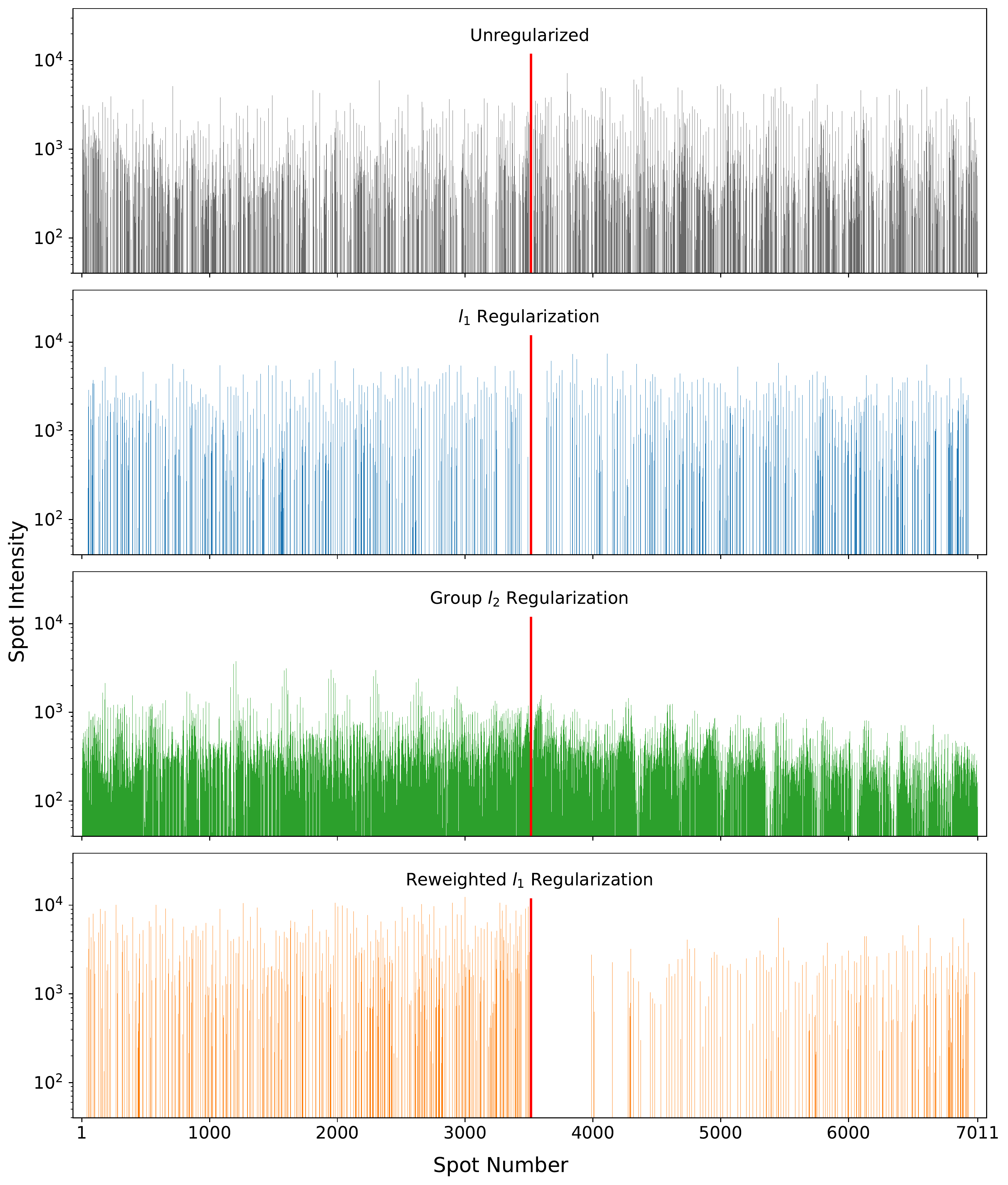}
	\caption{Optimal spot intensities resulting from the unregularized model and the $l_1$, group $l_2$, and reweighted $l_1$ regularized models ($\lambda = 5$) for patient $2$. The vertical red line divides the spots associated with the first beam (1--3516) from the second beam (3517--7011).}
	\label{fig:spot_vals}
\end{figure}

We first compared the results of the regularization methods on a single patient. Figure \ref{fig:spot_vals} depicts optimal spot intensities of the unregularized model and the $l_1$, group $l_2$, and reweighted $l_1$ regularized models for patient $2$. For all three regularizers, a regularization weight of $\lambda = 5$ was used; this choice accentuated the difference between their spot vectors. Without regularization, about one third of the total $7011$ spots are nonzero, with individual spot intensities ranging between $10^3$ and $10^4$. Under $l_1$ regularization, that fraction is reduced to only $13\%$, or $918$ nonzero spots, as the $l_1$ penalty encourages further sparsity. By contrast, with group $l_2$ regularization, the number of active spots increases significantly to $5099$ or $72\%$, while the average intensity drops to a little over $10^3$. Reweighted $l_1$ regularization produced a spot vector with the lowest number of nonzero elements: just $541$ or $7.7\%$ of the spots are nonzero. For patient $2$, these active spots tend to reside in the first beam, and their maximum intensity exceeds that of the other methods.

\begin{figure}[H]
	\centering
	\includegraphics[width=0.9\linewidth]{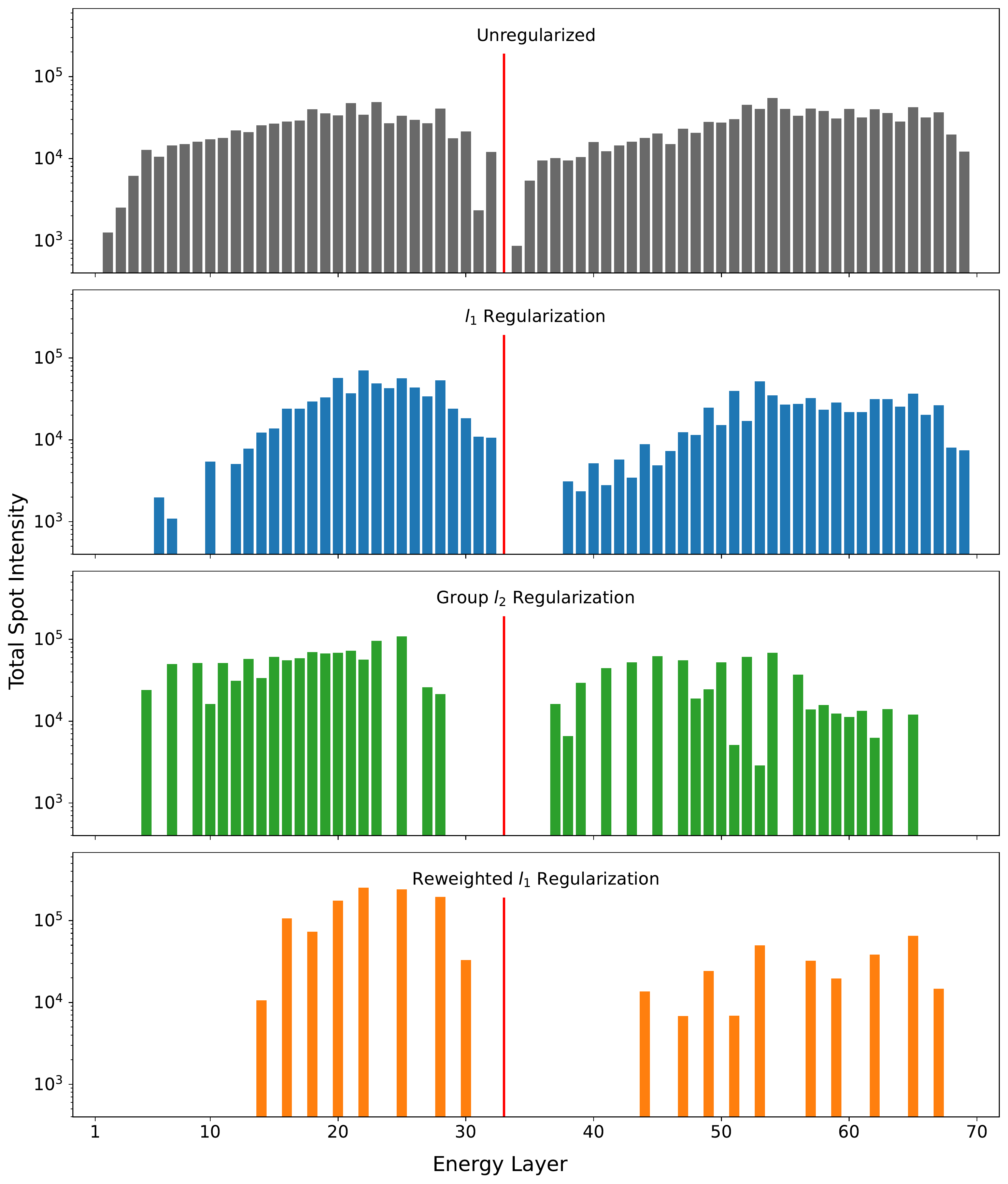}
	\caption{Sum of spot intensities in each energy layer (1--70) for the unregularized model and the $l_1$, group $l_2$, and reweighted $l_1$ regularized models ($\lambda = 5$) for patient $2$. The vertical red line divides the layers associated with the first beam (1--33) from the second beam (34--70).}
	\label{fig:layer_vals}
\end{figure}

Figure \ref{fig:layer_vals} shows the optimal intensity of the energy layers for patient $2$, using the three regularization models, all with $\lambda = 5$. Over $95\%$ of the total $70$ energy layers are nonzero under the unregularized model. These active layers are divided fairly evenly into two clusters, which coincide with the two beams delineated by the vertical red line. The total intensity of each energy layer averages between $10^4$ and $10^5$. With $l_1$ regularization, the fraction of nonzero energy layers drops to a modest $80\%$, where most of that reduction comes from deactivated layers at the edges of the clusters. Group $l_2$ regularization results in a steeper drop in the fraction of active energy layers, down to $61\%$ with additional sparsity in the middle of both beam clusters. However, the reweighted $l_1$ method performs better than both of these methods, cutting the number of nonzero energy layers down to only $18$ -- a reduction of over $75\%$ -- with a commensurate increase in the intensity of the active layers.

\begin{figure}[H]
	\centering
	\includegraphics[width=0.9\linewidth]{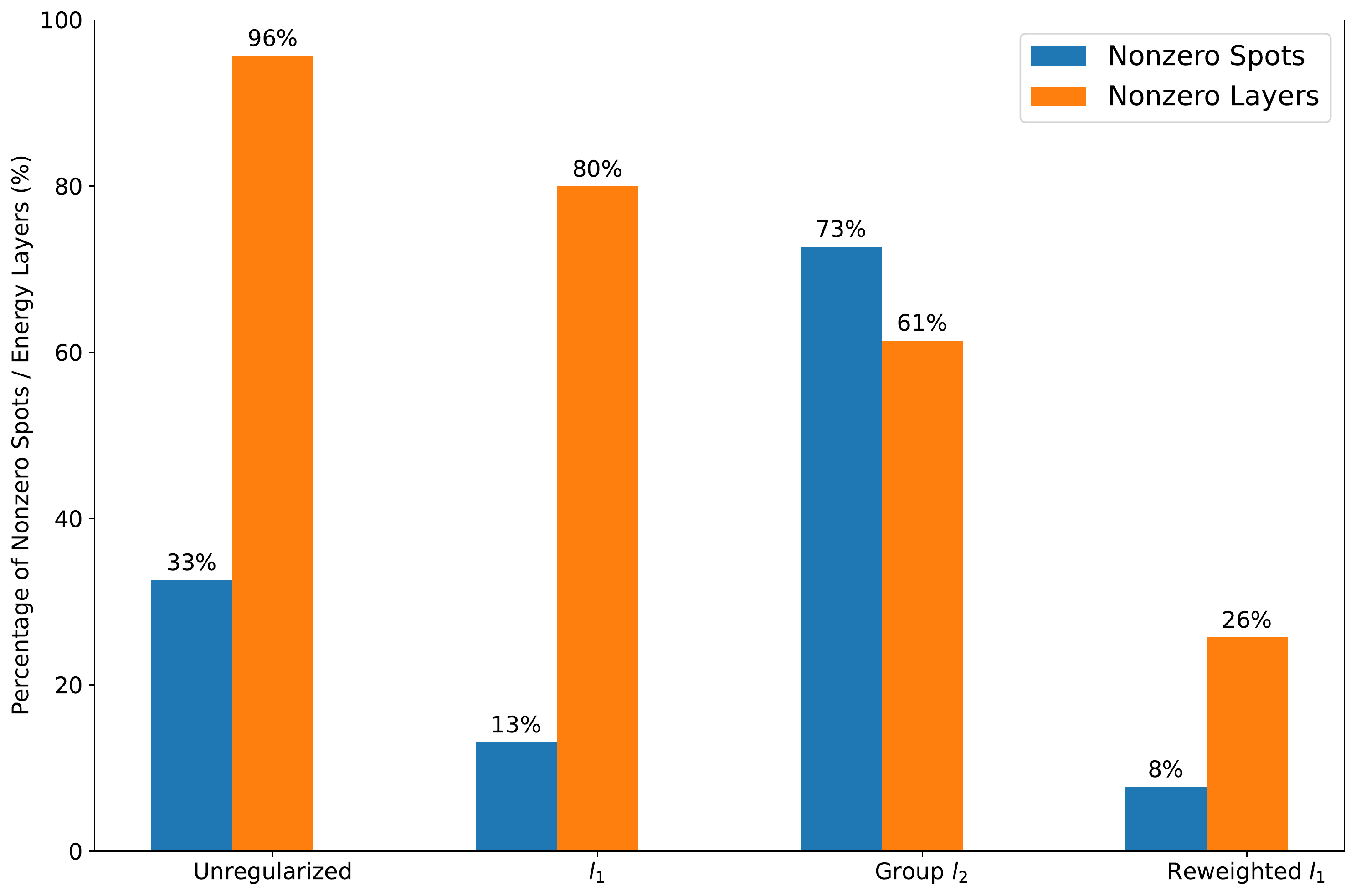}
	\caption{Percentage of nonzero spots/energy layers (relative to the total number of spots/layers) for the unregularized model and the $l_1$, group $l_2$, and reweighted $l_1$ regularized models ($\lambda = 5$) for patient $2$.}
	\label{fig:nnz_spots_layers}
\end{figure}

A summary of the results from the different regularization methods is given in Figure \ref{fig:nnz_spots_layers}. For a fixed $\lambda$, it is clear that reweighted $l_1$ achieves the lowest number of nonzero spots and nonzero energy layers out of all the methods.

\subsection{Trade-off between delivery efficiency and PTV coverage}
\label{sec:res_ptv}

\begin{figure}[H]
	\centering
	\includegraphics[width=0.9\linewidth]{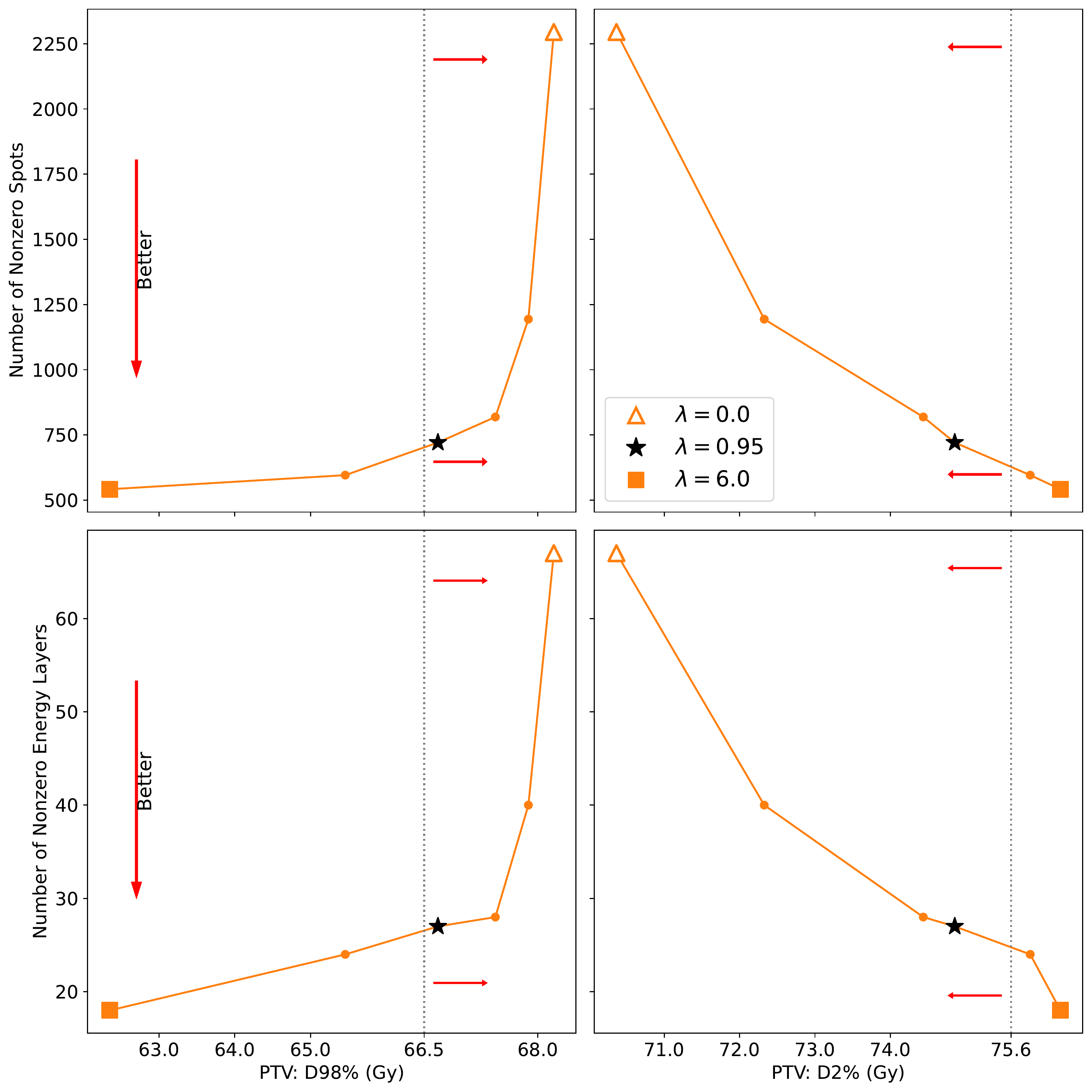}
	\caption{Number of nonzero spots (top) and energy layers (bottom) versus PTV dose percentile (left: D$98\%$; right: D$2\%$) for various values of the regularization weight $\lambda$, computed using the reweighted $l_1$ method on patient $2$. As $\lambda$ increases, the curves sweep from the $\triangle$ marker ($\lambda = 0.0$) to the $\blacksquare$ marker ($\lambda = 6.0$). The vertical gray dotted lines indicate clinical dose constraints on the PTV (D$98\% > 0.95p$ and D$2\% < 1.08p$, where $p = 70$ Gy is the prescription), and the red arrows indicate the directions of desirable change (increasing D$98\%$, decreasing D$2\%$, and decreasing number of nonzero spots/energy layers). A choice of weight $\lambda = 0.95$, marked by the $\bigstar$, produces a plan with good sparsity that respects the clinical constraints.}
	\label{fig:pareto_ptv}
\end{figure}

The regularization weight in the previous section was chosen to highlight the distinctions between the optimal intensity plots. However, $\lambda$ must be carefully selected to balance the trade-off between the total delivery time (highly correlated with the sparsity of the spots/energy layers) and the quality of the resulting treatment plan. Figure \ref{fig:pareto_ptv} examines this trade-off for reweighted $l_1$ regularization on patient $2$ using two measures of plan quality: D$98\%$ and D$2\%$ for the PTV. For different values of $\lambda$, we solved problem \ref{prob:reg} using the reweighted $l_1$ method, counted up the number of nonzero spots/energy layers, and calculated the optimal dose vector and PTV dose percentiles. We then plotted a point corresponding to this result in each of the subfigures of Figure \ref{fig:pareto_ptv} with the sparsity metric on the vertical axis and the plan quality metric on the horizontal axis (e.g., the unregularized point $\lambda = 0$ is marked by a triangle $\triangle$). By connecting the points in each subfigure, we obtained a set of Pareto optimal curves, which show the trade-off between plan delivery efficiency and plan quality.

The top left subfigure depicts the number of nonzero spots versus D$98\%$ to the PTV for $\lambda$ ranging from zero to $6.0$ (marked by the square). As $\lambda$ increases, the number of active spots decreases, but so does D$98\%$. A choice of $\lambda = 0.95$ (marked by the star) achieves the lowest number of nonzero spots $\approx 720$, while still maintaining D$98\%$ above $95\%$ of the prescription, indicated by the vertical gray dotted line at $66.5$ Gy. A similar plot can be seen in the bottom left subfigure, which shows the number of nonzero energy layers versus D$98\%$ to the PTV; the same choice of $\lambda$ yields $27$ active layers. On the righthand side, the subfigures display the number of nonzero spots (top) and energy layers (bottom) versus D$2\%$ to the PTV. As the regularization weight increases, D$2\%$ also increases, but never exceeds $108\%$ of the prescription (as indicated by the dotted line at $75.6$ Gy) for any $\lambda \leq 0.95$. Thus, out of all the weights, $\lambda = 0.95$ yields a good trade-off between sparsity and PTV coverage: it achieves a reduction of $89\%$ and $61\%$ in the number of spots and energy layers, respectively, while still fulfilling all target dose constraints.

\subsection{Trade-off between delivery efficiency and overall plan quality}
\label{sec:res_cost}

This section studies the Pareto optimal trade-off curves between spot/energy layer sparsity and treatment plan quality using different regularizers to determine which regularization method provides the {\em best} trade-off, i.e., the largest increase in sparsity for the least decrease in plan quality. Rather than plotting multiple dose-volume metrics, we focus on a single consolidated quality measure: the plan cost function (equation \ref{eq:cost}), which includes dose fidelity terms for the PTV and all organs-at-risk (OARs). A lower value of $f(\overline d, \underline d)$ at the optimum implies a higher quality treatment plan.

To facilitate comparison, we also focus on the {\em relative} change in sparsity (number of nonzero spots/energy layers) and plan cost with respect to the unregularized solution. Let $x_{unreg}$ be the optimal spots resulting from the unregularized problem \ref{prob:unreg}, and $x_{reg}$ be the optimal spots resulting from a particular regularization method. We evaluate the plan cost function given in equation \ref{eq:cost} on $x_{unreg}$ (with $\overline d_{unreg} = \max(Ax_{unreg} - p, 0)$ and $\underline d_{unreg} = -\min(Ax_{unreg} - p, 0)$) to obtain $c_{unreg}$, and similarly on $x_{reg}$ to obtain $c_{reg}$. Then, the relative percentage change in plan cost for the regularizer is $100(c_{reg} - c_{unreg})/c_{unreg}$. The relative change in the number of nonzero spots ($s$) and number of nonzero energy layers ($l$) is defined in a similar fashion as $100(s_{reg} - s_{unreg})/s_{unreg}$ and $100(l_{reg} - l_{unreg})/l_{unreg}$, respectively. Thus, to construct the trade-off curve, we solve the regularized problem for various values of $\lambda$ and plot the relative change in sparsity versus the relative change in plan cost at each solution point.

\begin{figure}[H]
	\centering
	\includegraphics[width=0.9\linewidth]{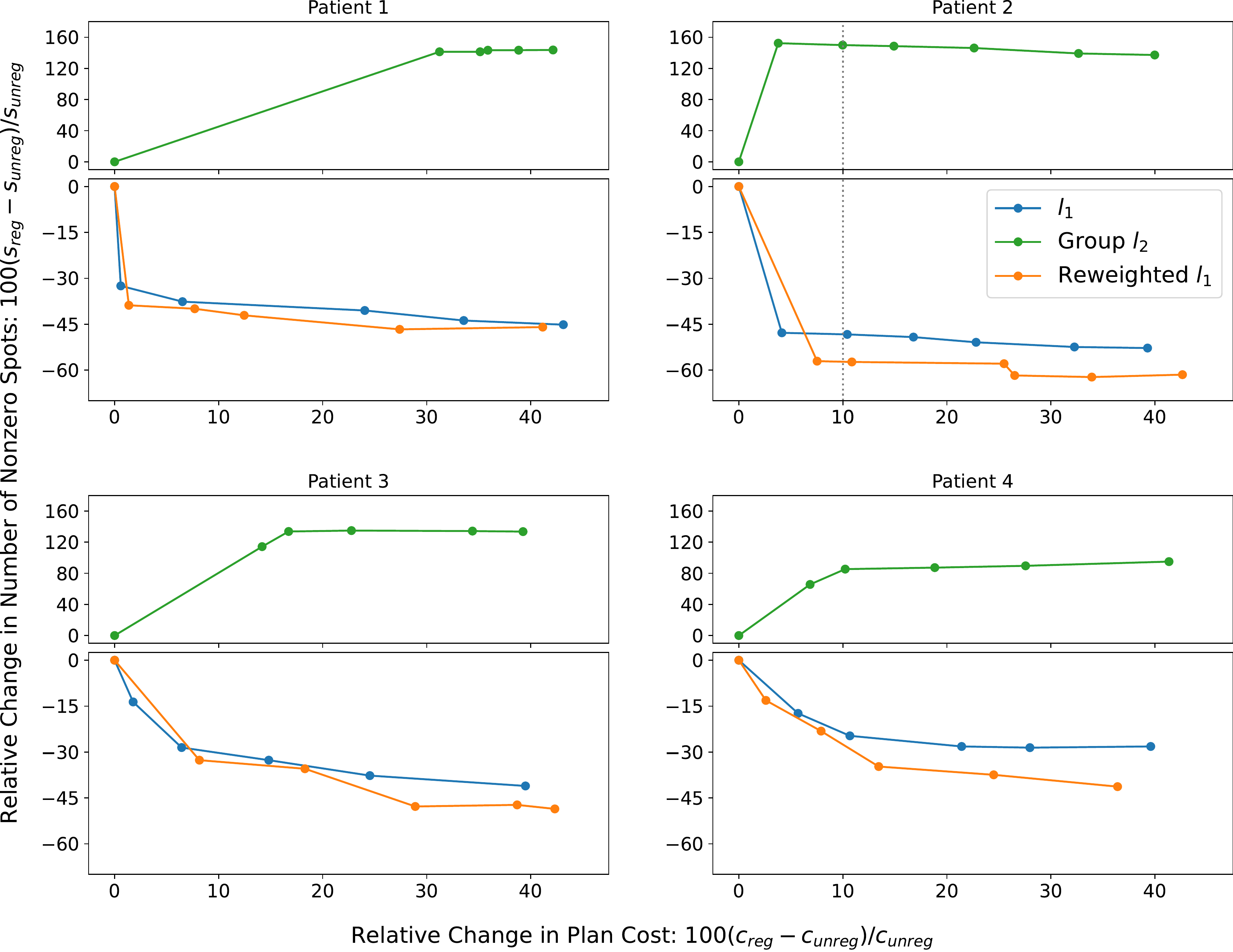}
	\caption{Relative change in number of nonzero spots versus relative cost to plan quality with respect to the unregularized model. For patient $2$, reweighted $l_1$ regularization achieves a $57\%$ reduction in the number of nonzero spots at only a $10\%$ cost to overall plan quality, relative to the unregularized model, as indicated by the vertical gray dotted line.} 
	\label{fig:pareto_spots_cost}
\end{figure}

For every patient, Figure \ref{fig:pareto_spots_cost} depicts the relative percentage change in the number of nonzero spots versus the relative percentage change in plan cost for the $l_1$, group $l_2$, and reweighted $l_1$ regularization methods. The origin corresponds to the unregularized plan ($\lambda = 0$). Both the $l_1$ and reweighted $l_1$ trade-off curves drop sharply from the origin, attaining on average a $30\%$ to $45\%$ decrease in nonzero spots for a less than $10\%$ increase in plan cost, with reweighted $l_1$ slightly outperforming $l_1$ by on average $5$ percentage points over all four patients. By contrast, the number of nonzero spots rises with group $l_2$ regularization, increasing up to $140\%$ within the first $10\%$ to $15\%$ increase in plan cost for all except patient $4$. This is consistent with our spot intensity plot for patient $2$ (Figure \ref{fig:spot_vals}), which shows the spot distribution is denser under group $l_2$ than without regularization.

\begin{figure}[H]
	\centering
	\includegraphics[width=0.9\linewidth]{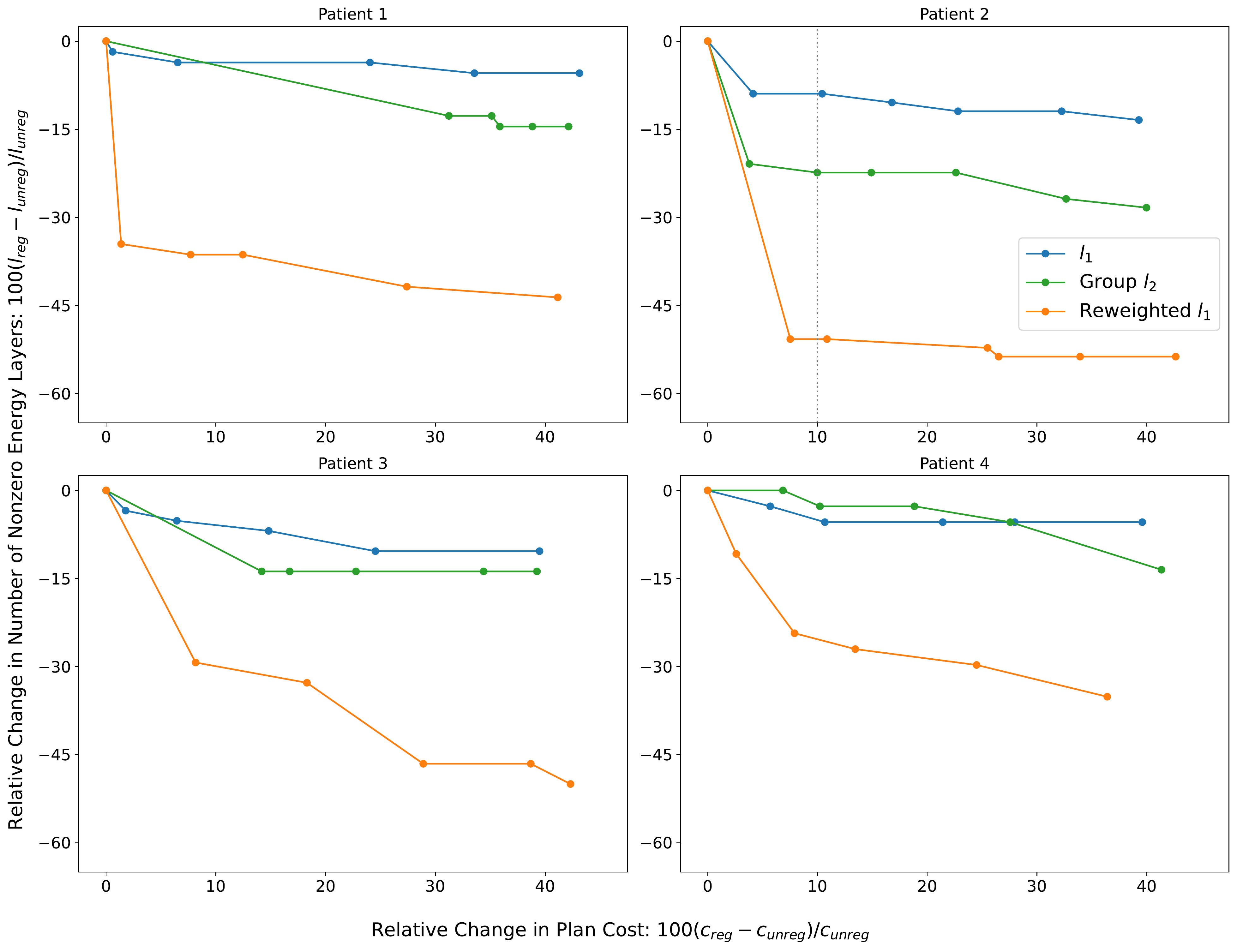}
	\caption{Relative change in number of nonzero energy layers versus relative cost to plan quality with respect to the unregularized model. For patient $2$, reweighted $l_1$ regularization achieves a $50\%$ reduction in the number of nonzero layers at only a $10\%$ cost to overall plan quality, relative to the unregularized model, as indicated by the vertical gray dotted line.}
	\label{fig:pareto_layers_cost}
\end{figure}

Figure \ref{fig:pareto_layers_cost} depicts the relative percentage change in the number of nonzero energy layers versus the relative percentage change in plan cost for the three regularization methods. Both $l_1$ and group $l_2$ trade-off curves decrease moderately from the origin, with group $l_2$ averaging about $9.5\%$ lower number of nonzero layers for a given percentage increase in cost. This matches our observations in Figures \ref{fig:layer_vals} and \ref{fig:nnz_spots_layers} that the group $l_2$ function is more effective at penalizing energy layers than the $l_1$ norm.

However, the reweighted $l_1$ method significantly outperforms both these regularizers. For patient $2$, it achieves an over $50\%$ decrease in the number of nonzero energy layers for a less than $10\%$ increase in plan cost. For the other patients, it provides a $25\%$ to $35\%$ reduction in active energy layers with a less than $15\%$ increment in plan cost. The average reduction in the number of nonzero layers from reweighted $l_1$ exceeds the best reduction from group $l_2$ by $12$ percentage points, and the majority of this reduction is realized with only about $10\%$ cost to treatment plan quality, relative to the unregularized plan.

\begin{figure}[H]
	\centering
	\includegraphics[width=0.9\linewidth]{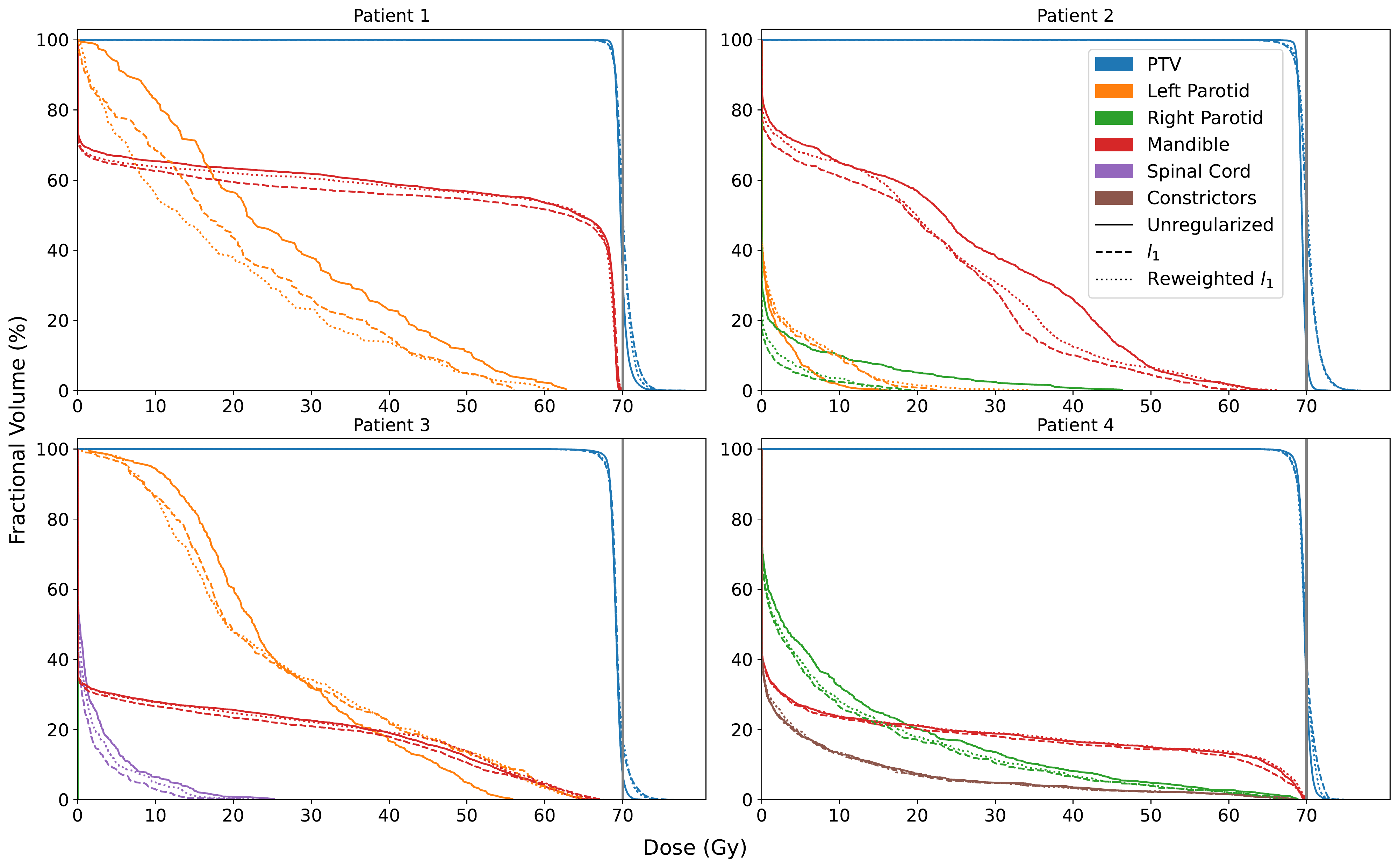}
	\caption{DVH curves for each patient obtained from the unregularized model (solid), and the standard $l_1$ (dashed) and reweighted $l_1$ (dotted) models regularized to approximately $10\%$ relative cost to plan quality. The vertical gray line indicates the prescription $p = 70$ Gy.}
	\label{fig:dvh_comp_cost}
\end{figure}

The vertical dotted lines in Figures \ref{fig:pareto_spots_cost} and \ref{fig:pareto_layers_cost} for patient $2$ correspond to a $10\%$ increase in the plan cost. The intersection of these lines with the Pareto curves of different regularization methods demonstrates the reduction in the number of nonzero spots and energy layers obtained using different regularizers. Figure \ref{fig:dvh_comp_cost} (top right) depicts the dose-volume histogram (DVH) curves of the unregularized plan, the $l_1$ regularized plan, and the reweighted $l_1$ regularized plan at the same $10\%$ relative change in plan cost. (We chose not to show the DVHs of group $l_2$ regularized plans because they overlap considerably with the DVHs of corresponding $l_1$ regularized plans, and as we saw earlier, group $l_2$ results in far worse delivery efficiency than $l_1$ or reweighted $l_1$ regularization). Compared to no regularization, the reweighted $l_1$ method reduces the number of active spots and energy layers by more than $50\%$, while providing relatively similar DVH curves with different trade-offs (compromised left parotid and PTV coverage/homogeneity, and improved right parotid and mandible). One can re-adjust the PTV/OAR weights in the plan cost function of the reweighted $l_1$ problem to achieve more uniform trade-offs. In the same vein, compared to standard $l_1$ regularization, reweighted $l_1$ reduces the number of active spots and energy layers by about $10\%$ and $40\%$, respectively, while producing almost identical DVH curves. The Pareto curves and DVHs of the other patients, also plotted at roughly $10\%$ relative change in plan cost, tell a similar story.

\subsection{Relative improvement in delivery time}
\label{sec:res_time}


The total plan delivery time is dependent on a multitude of machine-specific factors. In this section, we provide an estimate of how regularization directly impacts the delivery time, assuming a specific set of machine parameters. The total delivery time ($T$) can be approximated by
\begin{equation}
\label{eq:delivery_time}
	T = T_b \times (h_b(x) - 1)_+ + T_e \times (h_e(x) - 1)_+ + \sum_{g=1}^G T_s \times (h_s(x, \mathcal{I}_g) - 1)_+ + \frac{\sum_{j=1}^n x_j}{T_d},
\end{equation}
where $h_b(x)$ is the number of nonzero beams, $h_e(x)$ is the number of nonzero energy layers, $h_s(x, \mathcal{I}_g)$ is the number of nonzero spots in energy layer $g$, $T_b$ is the beam switching time (gantry rotation plus beam setup time), $T_e$ is the energy layer switching time, $T_s$ is the spot travel time, and $T_d$ is the proton dose rate \cite{GaoLin:2020,ZhangShenGao:2022}. Following van de Water et al. (2015) \cite{WaterHoogeman:2015}, we let $T_b = 30$ seconds, $T_e = 2$ seconds, $T_s = 0.01$ seconds, and $T_d = \frac{4 \times 10^{11}}{60}$ protons/second. We compute $T$ using the unregularized model and the $l_1$, group $l_2$, and reweighted $l_1$ regularized models, with regularization weight $\lambda$ chosen such that each regularized plan achieves a $10\%$ cost to plan quality. Then, we plot the relative percentage change in delivery time of each regularized model with respect to the unregularized model (i.e., $100(T_{reg} - T_{unreg})/T_{unreg}$).

\begin{figure}[H]
	\centering
	\includegraphics[width=0.9\linewidth]{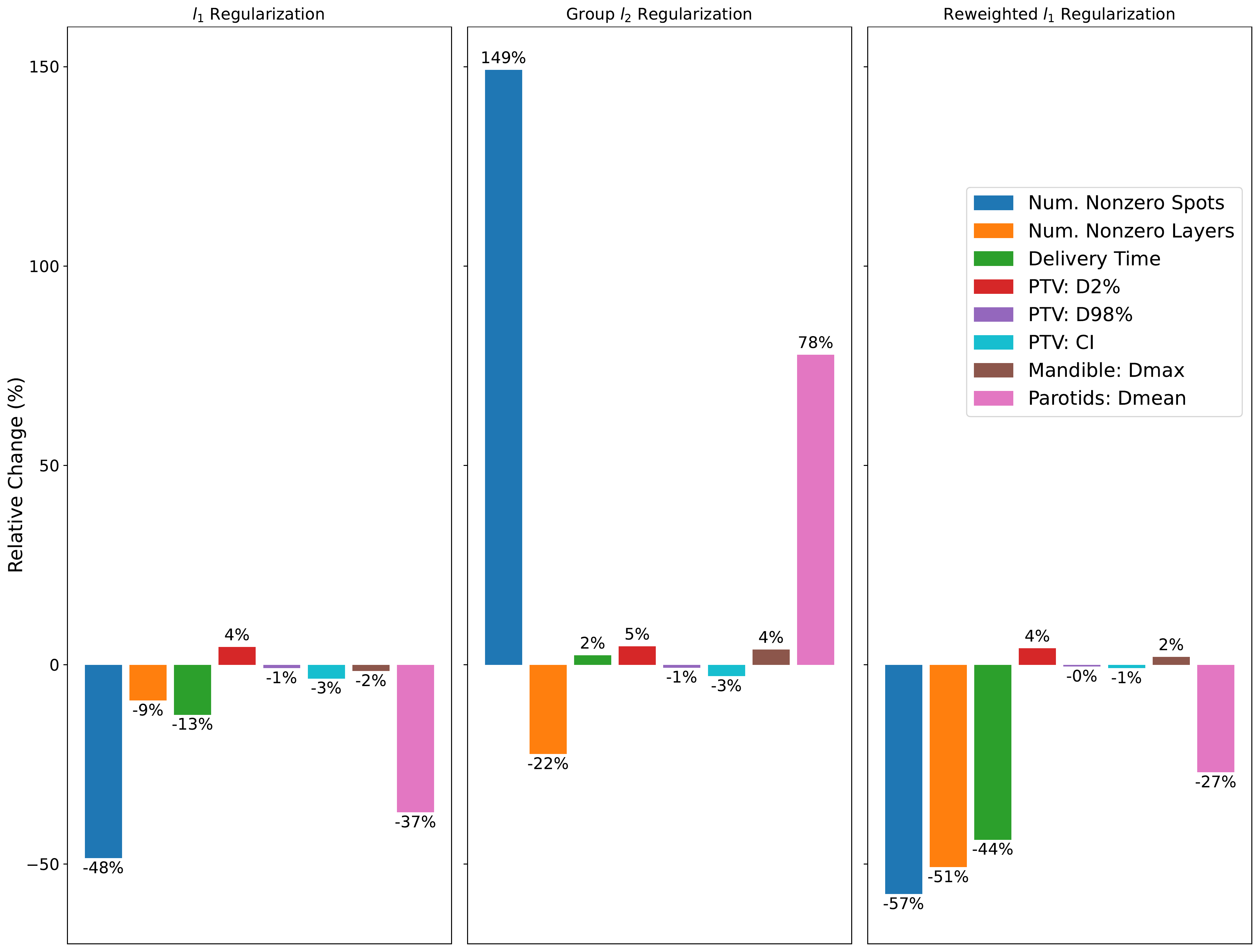}
	\caption{Relative change in delivery time and other plan metrics with respect to the unregularized model for $l_1$, group $l_2$, and reweighted $l_1$ regularization for patient $2$. For each regularizer, $\lambda$ was chosen such that the regularized model resulted in about $10\%$ cost to plan quality. Here the conformity index (CI) is defined as the total number of voxels that received at least $95\%$ of the prescribed dose, divided by the number of voxels in the PTV \cite{Petrova:2017}.}
	\label{fig:rel_change}
\end{figure}

The results for patient $2$ are shown in Figure \ref{fig:rel_change}. Standard $l_1$ reduces delivery time by a modest $13\%$, while group $l_2$ actually raises delivery time by $2\%$ due to the $149\%$ increase in the number of nonzero spots, which increases the spot delivery and spot travel time. By contrast, reweighted $l_1$ achieves a $44\%$ reduction in delivery time through its simultaneous reduction of the number of nonzero spots and nonzero energy layers by $57\%$ and $51\%$, respectively. This reduction comes at only a minor cost to the PTV and mandible -- D$2\%$ to the PTV goes up by $4\%$ and maximum dose to the mandible goes up by $2\%$. Results for the other patients reflect similar outcomes, with reweighted $l_1$ reducing total delivery time by between $20\%$ and $30\%$; see Table \ref{table:abs_metrics} and Figure \ref{fig:rel_change_all} in appendix \ref{app:plan_metrics} for additional data and plots.

\section{Discussion}
\label{sec:discussion}

This study proposed a method to improve the delivery efficiency of pencil beam scanning proton plans by simultaneously reducing the number of spots and energy layers using reweighted $l_1$ regularization. One can exactly model the spot/energy layer reduction problem using the $l_0$ regularizer, which in principle would improve plan delivery at the smallest possible cost to plan quality, but the $l_0$-regularized optimization problem is nonconvex and computationally prohibitive to solve. In imaging science and statistics, researchers often employ the $l_1$ norm as a convex surrogate for the $l_0$ norm, and in some cases (e.g., compressed sensing), the $l_1$ norm has proven to be just as effective as the $l_0$ norm at promoting sparsity \cite{CandesRombergTao:2006}. The reweighted $l_1$ regularization method was proposed \cite{CandesWakinBoyd:2008} to bridge the gap between the $l_0$ regularizer and the $l_1$ regularizer by better approximating the $l_0$ norm, while retaining the convexity of the $l_1$ norm. In proton treatment planning, this property translates to improving the plan delivery efficiency at a lower cost to plan quality, which we have demonstrated in this work. Our limited computational experiments on four head-and-neck cancer patients show that, for the same cost to plan quality, the reweighted $l_1$ method reduced the number of nonzero spots by up to $10$ percentage points more than standard $l_1$ and the number of nonzero energy layers by $25$ to $30$ percentage points more than group $l_2$ regularization.

Promoting spot/energy layer sparsity to improve plan delivery in IMPT is analogous to promoting beam profile smoothness to improve plan delivery in IMRT. Prior research has shown that plan delivery efficiency in IMRT can be significantly improved at minimal cost to dosimetric plan quality due to the phenomenon of {\em degeneracy} \cite{AlberReemtsen:2002,Llacer:2004}. The structure of the treatment planning problem results in a multitude of feasible plans with near-equal objective value (i.e., quality). This same phenomenon has been observed in IMPT planning problems \cite{WaterLomax:2020,WaterHoogeman:2015,GaoLin:2020,LinClasieGao:2019}, although unlike IMRT, it currently lacks a rigorous mathematical analysis. Our computational experiments demonstrated that with the reweighted $l_1$ method, one can reduce the number of spots and energy layers by on average $40\%$ and $35\%$, respectively, without significantly compromising the dosimetric plan quality.

In this study, we have adopted a constrained optimization framework, where the dosimetric plan quality is represented by a quadratic term in the objective, the sparsity promotion is carried out via a regularization penalty term, and the mean/max clinical dose criteria are enforced by hard constraints. However, the proposed reweighted $l_1$ method is agnostic to the optimization framework and can also be used in conjunction with an automation tool (e.g., hierarchical optimization \cite{ZarepishehDeasy:2022,Breedveld:2009,TaastiZarepisheh:2020}, multiple criteria optimization (MCO) \cite{CraftBortfeld:2008,MonzThieke:2008}, knowledge-based planning (KBP) \cite{AppenzollerMoore:2012,ShenNguyenJia:2020}). DVH constraints and plan robustness may be integrated into the optimization problem using existing techniques in the literature \cite{FuBoyd:2019,ZarepishehZinchenko:2013,MukherjeeZarepisheh:2020,robust_proton,TaastiZarepisheh:2020}. To limit the scope of this paper, we have not incorporated robustness into our formulation of the treatment planning problem. A preliminary robustness analysis of $13$ range and setup uncertainty scenarios shows that the reweighted $l_1$ method generates plans with a similar level of robustness to the unregularized model's plans. Moreover, even without accounting for these  uncertainties in the optimization, the DVH curves and clinical metrics of the plans lie within an acceptable range across the majority of scenarios. Appendix \ref{app:robustness} provides details on our analysis and the resulting figures.

Finally, we mention that in this proof-of-concept work, we have not enforced the machine-specific minimum-monitor-unit (min-MU) constraint. One can enforce the min-MU constraint using a two-step optimization method as described in Lin et al. (2019) \cite{LinClasieGao:2019}: the first step identifies the active spots/energy layers (i.e., those with intensity greater than a pre-determined value), and the second step removes the inactive spots/energy layers and enforces the min-MU constraint on the remaining spots. Increasing the min-MU threshold also allows for a higher dose rate, which can accelerate the delivery of each spot. This is especially important because the intensities of the active spots usually increase with the overall sparsity of the spots/energy layers in the treatment plan. Gao et al. (2020) \cite{GaoLin:2020} suggested using different min-MU thresholds for each energy layer to further increase the dose rate and expedite spot delivery.

\section{Conclusion}
\label{sec:conclusion}

The reweighted $l_1$ regularization method is capable of simultaneously reducing the number of spots and energy layers in a proton treatment plan, while imposing minimal cost to dosimetric plan quality. Moreover, it achieves a better trade-off between delivery efficiency and plan quality than standard $l_1$ and group $l_2$ regularization. Thus, reweighted $l_1$ regularization is a powerful method for improving the delivery of proton therapy.


\section*{Acknowledgments}

This work was partially supported by MSK Cancer Center Support Grant/Core Grant from the NIH (P30 CA008748).

\bibliographystyle{alpha}
\bibliography{reweighted_l1}
\nocite{BoydVandenberghe:2004}

\pagebreak

\appendix

\section{Derivation of problem \ref{prob:unreg}}
\label{app:prob_simp}

Problem \ref{prob:unreg_ncvx} in the main text is a nonconvex optimization problem because the equality constraints $\overline d = (Ax - p)_+$ and $\underline d = (Ax - p)_-$ are nonlinear. We will show that it is equivalent to the convex optimization problem \ref{prob:unreg}, which replaces the nonlinear equality constraints with the linear equality constraint $Ax - \overline d + \underline d = p$. Since the objective functions are identical, it is sufficient to prove that any solution of problem \ref{prob:unreg_ncvx} is a feasible point for problem \ref{prob:unreg} and vice versa.

Let $v^{\star} := (x^{\star}, \overline d^{\star}, \underline d^{\star})$ be a solution of problem \ref{prob:unreg_ncvx}. Then,
\begin{align*}
	Ax^{\star} - \overline d^{\star} + \underline d^{\star} - p
	&= (Ax^{\star} - p) - (Ax^{\star} - p)_+ + (Ax^{\star} - p)_- \\
	&= (Ax^{\star} - p) - (\max(Ax^{\star} - p, 0) + \min(Ax^{\star} - p, 0)) = 0,
\end{align*}
because any real function can be written as the sum of its nonnegative and nonpositive parts. Thus, $v^{\star}$ is feasible for problem \ref{prob:unreg}, and since the two problems have the same objective function, $v^{\star}$ is a solution of problem \ref{prob:unreg}.

Now let us show the converse. First, we will relax the equality constraints in problem \ref{prob:unreg_ncvx} to get
\begin{equation}
	\label{prob:unreg_relax}
	\begin{array}{ll}
		\mbox{minimize} & f(\overline d, \underline d) \\
		\mbox{subject to} & \overline d \geq (Ax - p)_+, \quad \underline d \geq (Ax - p)_-, \quad Bx \leq c \\
		& x \geq 0, \quad \overline d \geq 0, \quad \underline d \geq 0.
	\end{array}
\end{equation}
This relaxed problem is equivalent to problem \ref{prob:unreg_ncvx} because the relaxed constraints, $\overline d \geq (Ax - p)_+$ and $\underline d \geq (Ax - p)_-$, are tight at the optimum, since the objective function \ref{eq:cost} is monotonically increasing over the nonnegative reals. Any feasible point $(\hat x, \hat{\overline d}, \hat{\underline d})$ of problem \ref{prob:unreg_relax} that does not satisfy $\hat{\overline d} = (A\hat x - p)_+$ and $\hat{\underline d} = (A\hat x - p)_-$ cannot be optimal, as we can perturb it to obtain another feasible point with a strictly lower objective value.

For example, suppose $\hat{\underline d}_i > \min(A\hat x - p, 0)_i$ for a voxel $i \in \{1,\ldots,m\}$, then we can decrease $\hat{\underline d}_i$ by some small quantity $\delta > 0$ to get $\tilde{\underline d} := \hat{\underline d} - \delta e_i \geq 0$ that still satisfies the inequality constraint. (Here $e_i \in \{0,1\}^n$ is the indicator vector of index $i$, which equals 1 at $i$ and 0 elsewhere). The new point $(\hat x, \hat{\overline d}, \tilde{\underline d})$ is feasible for problem \ref{prob:unreg_relax}, but attains a lower objective $f(\hat{\overline d}, \tilde{\underline d}) < f(\hat{\overline d}, \hat{\underline d})$ because the quadratic term $\underline w_i\tilde{\underline d}_i^2 < \underline w_i\hat{\underline d}_i^2$ by our choice of $\delta > 0$, assuming $\underline w_i \neq 0$. If $\underline w_i = 0$, the underdose to voxel $i$ has no impact on the objective and can be removed entirely from the optimization problem.

It suffices then to show that any solution $v^{\star} = (x^{\star}, \overline d^{\star}, \underline d^{\star})$ of problem \ref{prob:unreg} is a solution of problem \ref{prob:unreg_relax}. This is true because $Ax^{\star} - \overline d^{\star} + \underline d^{\star} = p$ and $\overline d^{\star} \geq 0, \underline d^{\star} \geq 0$ imply
\begin{align*}
	\overline d^{\star} &= Ax^{\star} - p + \underline d^{\star} \geq Ax^{\star} - p, \\
	\underline d^{\star} &= -(Ax^{\star} - p) + \overline d^{\star} \geq -(Ax^{\star} - p).
\end{align*}
Thus, we can conclude that 
\begin{align*}
	\overline d^{\star} &\geq \max(Ax^{\star} - p, 0) = (Ax^{\star} - p)_+, \\ \underline d^{\star} &\geq \max(-(Ax^{\star} - p), 0) = -\min(Ax^{\star} - p, 0) = (Ax^{\star} - p, 0)_-.
\end{align*}


\section{Delivery efficiency and plan quality metrics}
\label{app:plan_metrics}

\begin{figure}[H]
	\centering
	\includegraphics[width=0.9\linewidth]{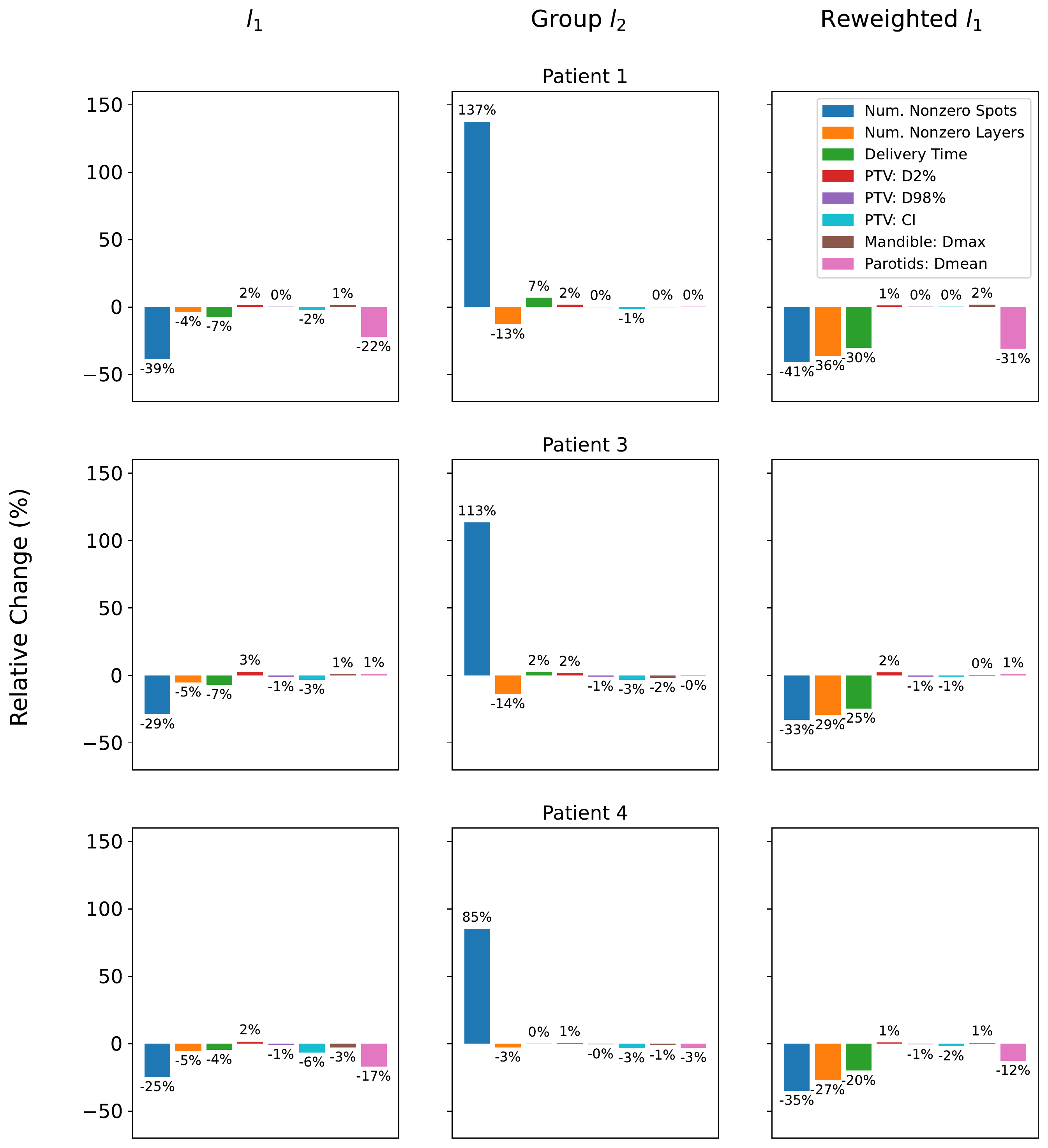}
	\caption{Relative change in delivery time and other plan metrics with respect to the unregularized model for $l_1$, group $l_2$, and reweighted $l_1$ regularization. For each regularizer, $\lambda$ was chosen such that the regularized model resulted in about $10\%$ cost to plan quality. Here the conformity index (CI) is defined as the total number of voxels that received at least $95\%$ of the prescribed dose, divided by the number of voxels in the PTV \cite{Petrova:2017}.}
	\label{fig:rel_change_all}
\end{figure}

In this section, we present further results on the delivery time and other clinical metrics. Figure \ref{fig:rel_change_all} is an extension of Figure \ref{fig:rel_change} in the main text, depicting the relative percentage change in total delivery time, number of nonzero spots/energy layers, and several dosimetric quantities at $10\%$ cost to plan quality for patients $1$, $3$, and $4$. Table \ref{table:abs_metrics} provides the absolute values for the metrics plotted in the two figures. Together, they show that reweighted $l_1$ regularization outperforms $l_1$ and group $l_2$ regularization in all patients.

\begin{table}
	\centering
	\begin{tabular}{l|l|rrrr}
		\hline \hline
		& & Unregularized & $l_1$ & Group $l_2$ & Reweighted $l_1$ \\
		\hline
		\multirow{8}{*}{Patient 1}
			& Num. Nonzero Spots   & 1796    & 1102    & 4264    & 1061  \\
			& Num. Nonzero Layers  &   55    &   53    &   48    &   35  \\
			& Delivery Time (s)    &  155.41 &  144.49 &  166.16 & 108.26 \\
			& PTV: D2\% (Gy)       &   71.90 &   73.10 &   73.18 & 72.70 \\
			& PTV: D98\% (Gy)      &   68.17 &   68.46 &   68.29 & 68.34 \\
			& PTV: CI              &    1.40 &    1.37 &    1.38 &  1.40 \\
			& Mandible: Dmax (Gy)  &   67.49 &   68.50 &   67.62 & 68.77 \\
			& Parotids: Dmean (Gy) &   11.50 &    8.96 &   11.53 &  7.95 \\
		\hline
		\multirow{8}{*}{Patient 2}
			& Num. Nonzero Spots   & 2295    & 1182    & 5719    & 976    \\
			& Num. Nonzero Layers  &   67    &   61    &   52    &  33    \\
			& Delivery Time (s)    &  184.28 &  161.21 &  188.67 & 103.43 \\
			& PTV: D2\% (Gy)       &   70.37 &   73.49 &   73.61 &  73.30 \\
			& PTV: D98\% (Gy)      &   68.21 &   67.62 &   67.73 &  67.93 \\
			& PTV: CI              &    1.49 &    1.44 &    1.45 &   1.48 \\
			& Mandible: Dmax (Gy)  &   64.86 &   63.84 &   67.36 &  66.13 \\
			& Parotids: Dmean (Gy) &    2.04 &    1.29 &    3.63 &   1.49 \\
		\hline
		\multirow{8}{*}{Patient 3}
			& Num. Nonzero Spots   & 1751    & 1247    & 3738    & 1172    \\
			& Num. Nonzero Layers  &   58    &   55    &   50    &   41    \\
			& Delivery Time (s)    &  160.93 &  149.92 &  164.88 &  121.31 \\
			& PTV: D2\% (Gy)       &   70.54 &   72.44 &   71.95 &   72.12 \\
			& PTV: D98\% (Gy)      &   67.70 &   66.97 &   67.06 &   67.14 \\
			& PTV: CI              &    1.84 &    1.78 &    1.78 &    1.83 \\
			& Mandible: Dmax (Gy)  &   67.06 &   67.53 &   65.91 &   67.11 \\
			& Parotids: Dmean (Gy) &   13.41 &   13.54 &   13.37 &   13.51 \\
		\hline
		\multirow{8}{*}{Patient 4}
			& Num. Nonzero Spots   & 259    & 195    & 480    & 169    \\
			& Num. Nonzero Layers  &  37    &  35    &  36    &  27    \\
			& Delivery Time (s)    & 104.22 &  99.60 & 104.44 &  83.42 \\
			& PTV: D2\% (Gy)       &  71.27 &  72.47 &  71.77 &  71.88 \\
			& PTV: D98\% (Gy)      &  68.28 &  67.71 &  67.95 &  67.88 \\
			& PTV: CI              &   1.59 &   1.48 &   1.54 &   1.56 \\
			& Mandible: Dmax (Gy)  &  64.03 &  62.29 &  63.32 &  64.55 \\
			& Parotids: Dmean (Gy) &   7.01 &   5.83 &   6.79 &   6.14 \\
		\hline \hline
	\end{tabular}
	\caption{Values of various metrics under the unregularized plan and the $l_1$, group $l_2$, and reweighted $l_1$ regularized plans. For each patient and regularizer, $\lambda$ was chosen such that the regularized plan resulted in a $10\%$ cost to plan quality relative to the unregularized plan (e.g., as shown by the dotted line in Figure \ref{fig:pareto_spots_cost}).}
	\label{table:abs_metrics}
\end{table}

\subsection{Robustness analysis}
\label{app:robustness}

To analyze the robustness of the proton plans, we generated $13$ uncertainty scenarios (including the nominal scenario) by rescaling the stopping power ratio (SPR) image $\pm 3.5\%$ to simulate range over/undershoot errors \cite{TaastiRichter:2018,Paganetti:2012} and shifting the isocenter $\pm 3$ mm in the $x$, $y$, and $z$ directions to simulate setup errors. For each scenario $r \in \{1,\ldots,13\}$, we computed the dose influence matrix $A^{r} \in \mathbf{R}_+^{m \times n}$. We solved the unregularized treatment planning problem \ref{prob:unreg} using only the dose influence matrix of the nominal scenario to get the optimal nominal spot vector $x_{unreg}^{\star}$ and calculated the dose in each scenario to be $d_{unreg}^{r} = A^rx_{unreg}^\star$. We repeated this process, again using only the nominal dose influence matrix in the optimization, with the reweighted $l_1$ regularization method to obtain doses $d_{rewl_1}^{r} = A^rx_{rewl_1}$. (As before, $\lambda$ was chosen such that the reweighted $l_1$ plan results in a $10\%$ cost to plan quality relative to the unregularized plan). In the end, for each patient, we had two sets of dose vectors representing the potential scenario outcomes if we were to treat using the unregularized plan and the reweighted $l_1$ regularized plan: $(d_{unreg}^1,\ldots,d_{unreg}^{13})$ and $(d_{rewl_1}^1,\ldots,d_{rewl_1}^{13})$.

\begin{figure}[H]
	\centering
	\includegraphics[width=0.9\linewidth]{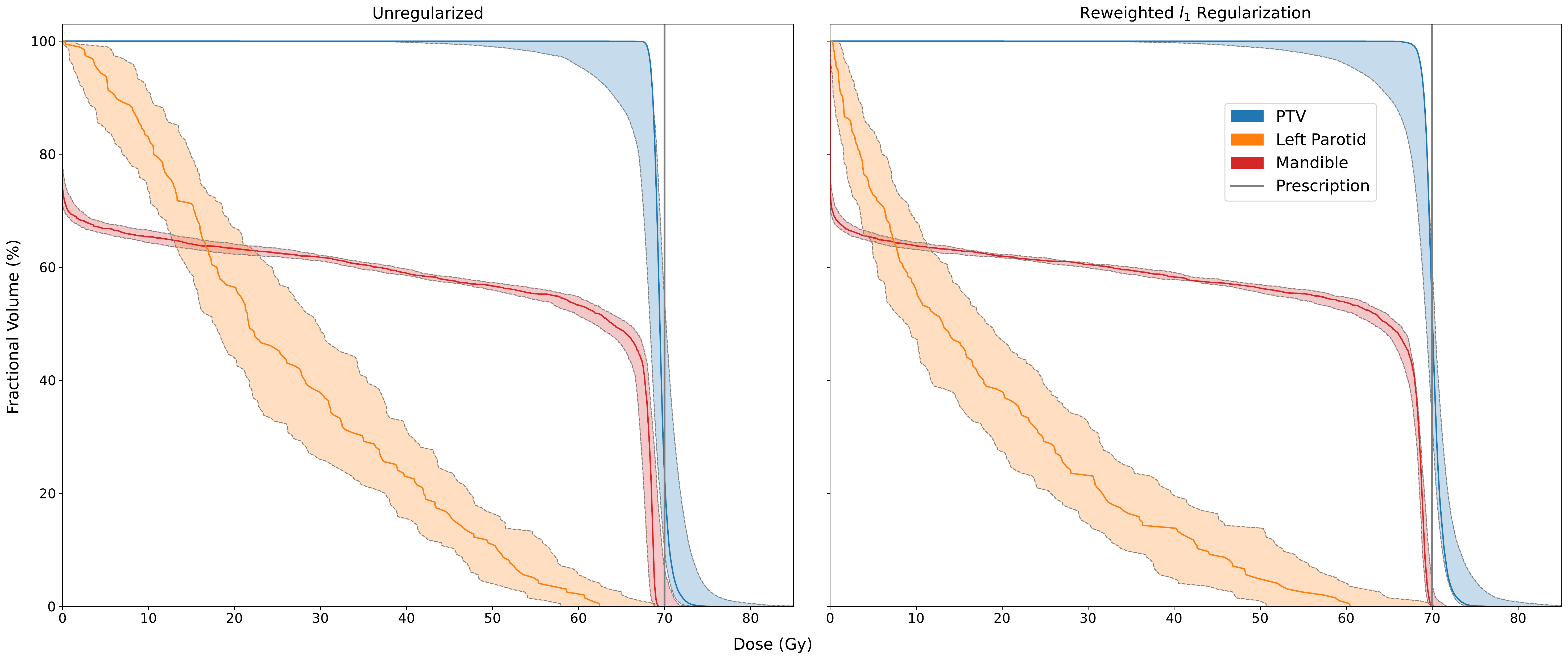}
	\caption{DVH bands across all uncertainty scenarios obtained from the unregularized plan and the reweighted $l_1$ regularized plan for patient $1$. The solid lines indicate the DVH curves of the nominal scenario.}
	\label{fig:dvh_bands-p1}
\end{figure}

\begin{figure}[H]
	\centering
	\includegraphics[width=0.9\linewidth]{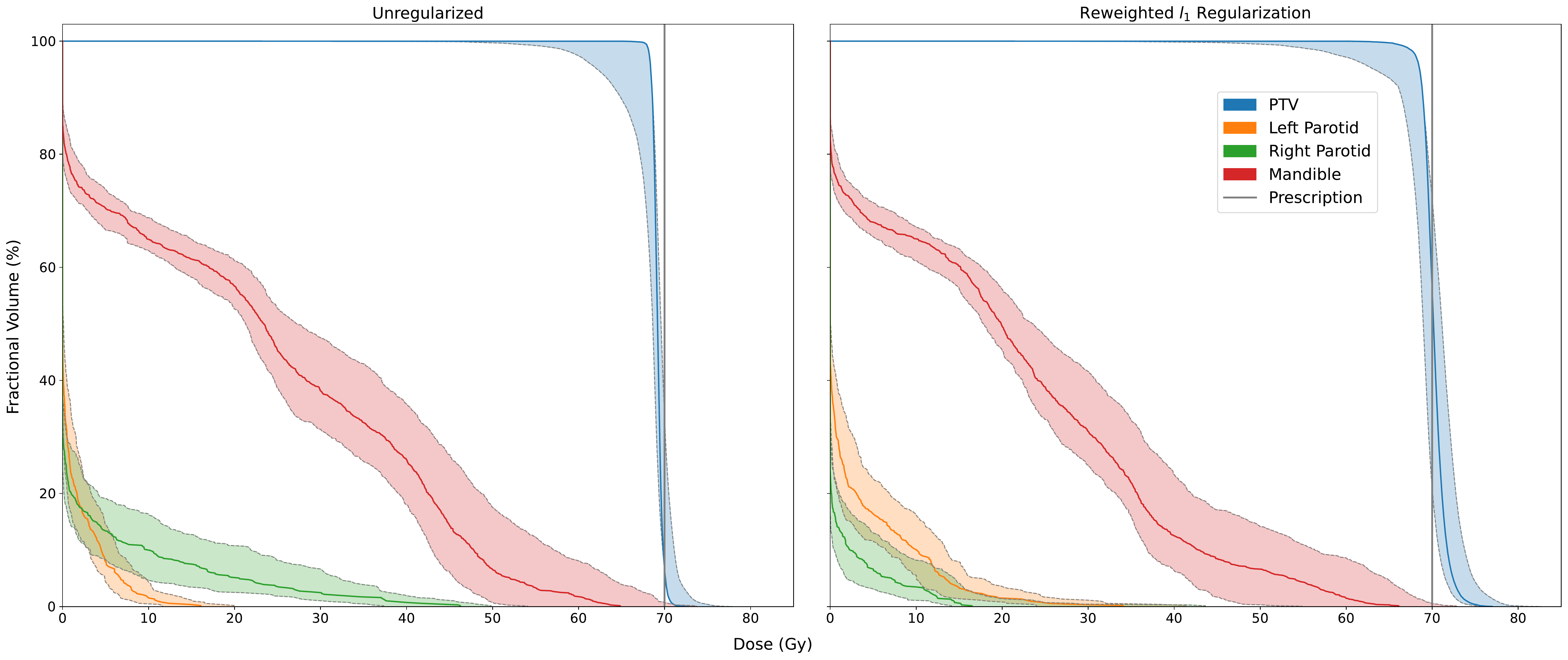}
	\caption{DVH bands across all uncertainty scenarios obtained from the unregularized plan and the reweighted $l_1$ regularized plan for patient $2$. The solid lines indicate the DVH curves of the nominal scenario.}
	\label{fig:dvh_bands-p2}
\end{figure}

\begin{figure}[H]
	\centering
	\includegraphics[width=0.9\linewidth]{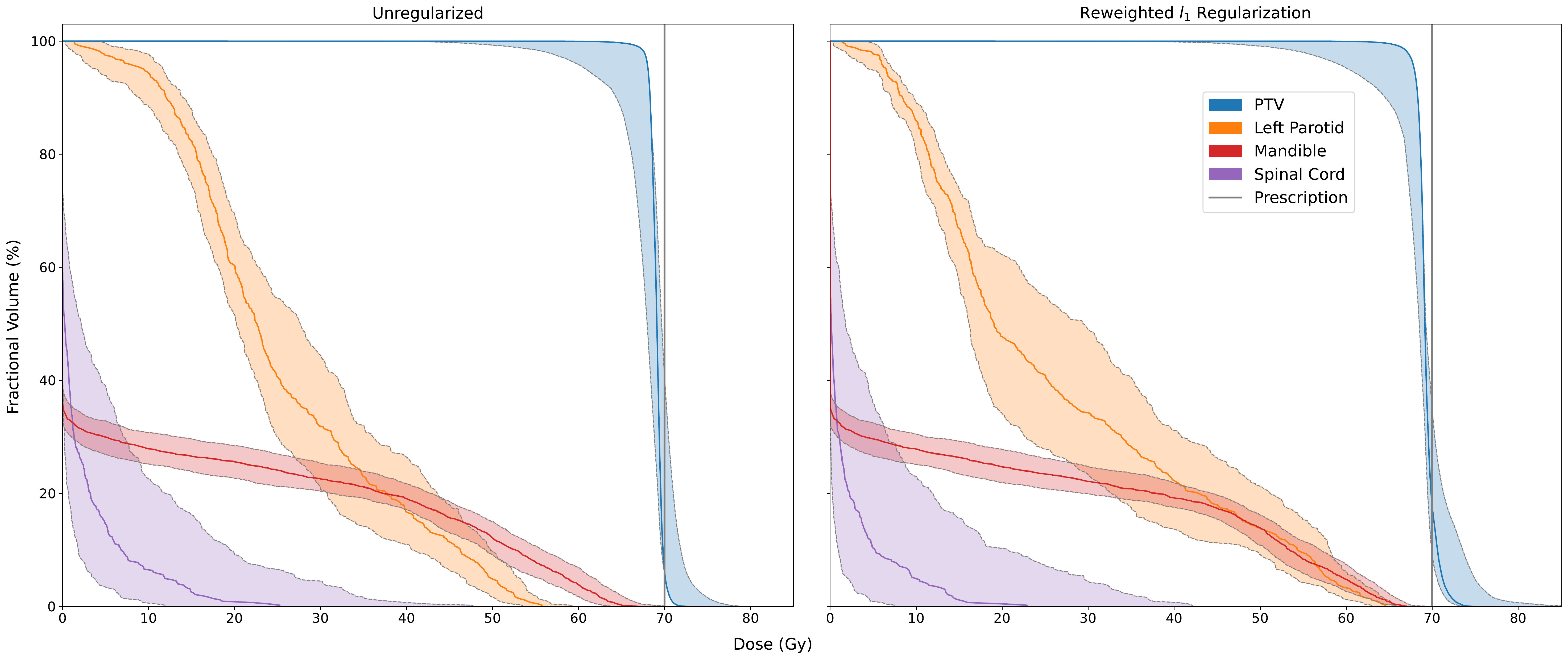}
	\caption{DVH bands across all uncertainty scenarios obtained from the unregularized plan and the reweighted $l_1$ regularized plan for patient $3$. The solid lines indicate the DVH curves of the nominal scenario.}
	\label{fig:dvh_bands-p3}
\end{figure}

\begin{figure}[H]
	\centering
	\includegraphics[width=0.9\linewidth]{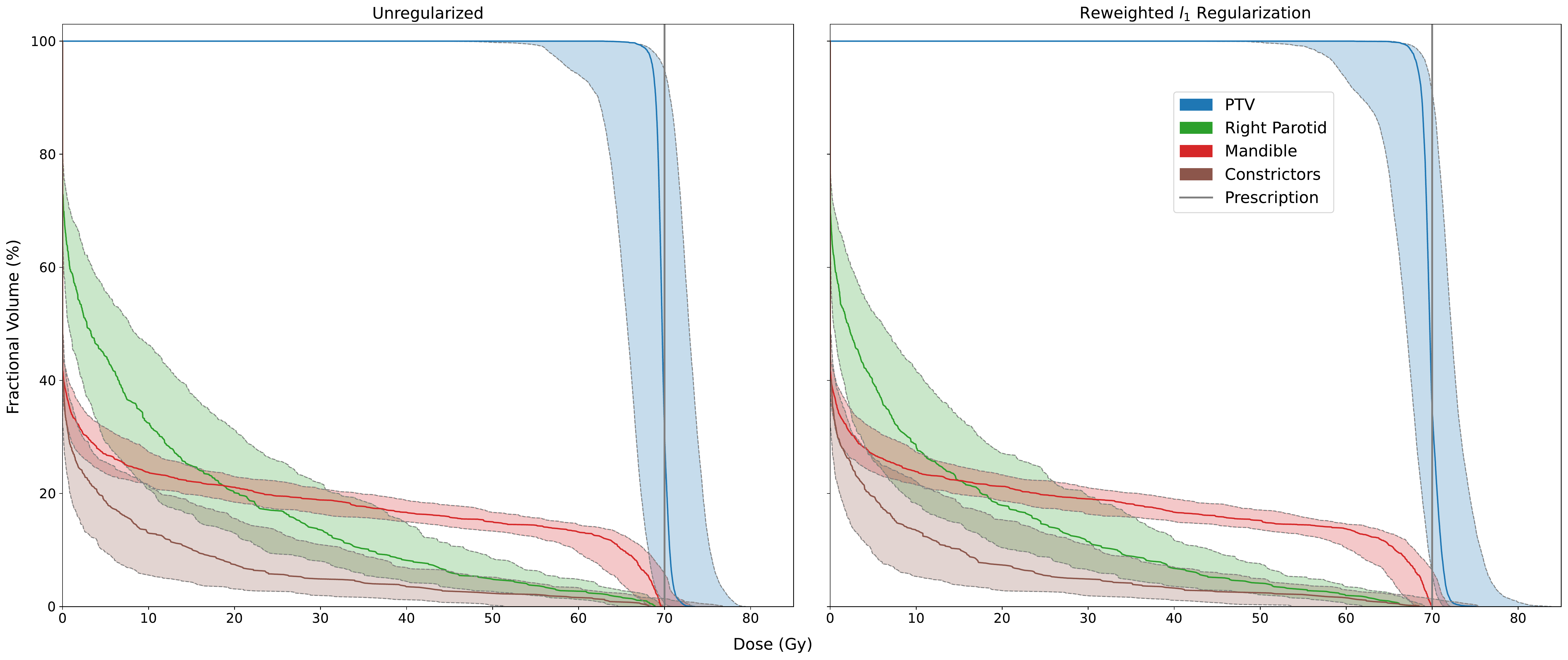}
	\caption{DVH bands across all uncertainty scenarios obtained from the unregularized plan and the reweighted $l_1$ regularized plan for patient $4$. The solid lines indicate the DVH curves of the nominal scenario.}
	\label{fig:dvh_bands-p4}
\end{figure}

Figures \ref{fig:dvh_bands-p1}--\ref{fig:dvh_bands-p4} depict the DVH bands resulting from these sets of doses for each patient. Every band delineates the range of DVH curves for a particular structure, taken across all uncertainty scenarios. The corresponding solid line is the DVH curve in the nominal scenario. Generally, the DVH bands of the reweighted $l_1$ regularized plan are similar to those of the unregularized plan. For patients $1$ and $4$, the reweighted $l_1$ plan results in a slightly narrower PTV band (smaller range of uncertainty), while the opposite is true for patients $2$ and $3$. In the case of OARs, the most noticeable difference is the reweighted $l_1$ plan exhibits thinner bands for the mandible in patient $1$, but wider bands for the left parotid in patient $3$. All other structures have comparable DVH bands between the unregularized and the reweighted $l_1$ regularized plans. We also provide the median and range of several clinical metrics in Table \ref{table:metric_robust}. These values support our conclusion that the reweighted $l_1$ method produces a treatment plan with similar robustness to the plan generated by the unregularized model, and its resulting dose satisfies clinical constraints in the majority of our uncertainty scenarios.

\begin{table}
	\centering
	\begin{tabular} {l|l|rr|rr}
		\hline \hline
		& & \multicolumn{2}{c|}{Unregularized} & \multicolumn{2}{c|}{Reweighted $l_1$} \\
		\cline{3-6}
		& & Median & Range & Median & Range \\
		\hline
		\multirow{4}{*}{Patient 1}
		& PTV: D2\% (Gy)           & 72.73 &  4.84 & 73.36 &  4.42 \\
		& PTV: D98\% (Gy)          & 64.06 & 12.69 & 62.96 & 13.60 \\
		& Mandible: Dmax (Gy)      & 69.58 &  4.55 & 70.91 &  1.80 \\
		& Left Parotid: Dmean (Gy) & 25.70 &  9.42 & 17.76 &  8.14 \\
		\hline
		\multirow{5}{*}{Patient 2}
		& PTV: D2\% (Gy)            & 71.03 &  2.48 & 73.58 &  3.97 \\
		& PTV: D98\% (Gy)           & 64.89 &  9.07 & 64.18 &  9.89 \\
		& Mandible: Dmax (Gy)       & 65.80 & 19.66 & 67.26 & 18.00 \\
		& Left Parotid: Dmean (Gy)  &  1.15 &  1.16 &  2.32 &  2.08 \\
		& Right Parotid: Dmean (Gy) &  2.71 &  3.22 &  0.87 &  1.55 \\
		\hline
		\multirow{5}{*}{Patient 3}
		& PTV: D2\% (Gy)            & 71.66 &  4.08 & 72.73 &  5.33 \\
		& PTV: D98\% (Gy)           & 64.30 & 11.23 & 62.54 & 12.01 \\
		& Mandible: Dmax (Gy)       & 67.33 &  5.63 & 67.12 &  3.31 \\
		& Left Parotid: Dmean (Gy)  & 25.61 &  6.28 & 25.80 &  9.45 \\
		& Spinal Cord: Dmax (Gy)    & 26.79 & 35.76 & 23.50 & 34.56 \\
		\hline
		\multirow{5}{*}{Patient 4}
		& PTV: D2\% (Gy)            & 71.29 &  7.97 & 71.85 &  7.62 \\
		& PTV: D98\% (Gy)           & 64.21 & 12.14 & 63.16 & 11.83 \\
		& Mandible: Dmax (Gy)       & 69.70 &  4.94 & 70.22 &  3.69 \\
		& Right Parotid: Dmean (Gy) & 10.92 &  8.98 &  9.65 &  8.27 \\
		& Constrictors: Dmean (Gy)  &  4.58 &  5.70 &  4.55 &  5.72 \\
		\hline \hline 
	\end{tabular}
	\caption{The median and range (maximum $-$ minimum) of various metrics calculated across $13$ uncertainty scenarios, using the unregularized plan and the reweighted $l_1$ regularized plan optimized with respect to the nominal scenario. For each patient, $\lambda$ was chosen such that the regularized plan resulted in a $10\%$ cost to plan quality relative to the unregularized plan.}
	\label{table:metric_robust}
\end{table}
	
\end{document}